\documentclass[a4paper,11pt]{article}
\pdfoutput=1

\usepackage{jinstpub}

\RequirePackage{subfig}
\RequirePackage{mhchem}
\RequirePackage{amssymb}
\RequirePackage{booktabs}
\RequirePackage{multirow}

\newcommand{\ttbar}{\ensuremath{t\bar{t}}}
\newcommand{\TeV}{\ensuremath{\text{Te\kern -0.1em V}}}
\newcommand{\GeV}{\ensuremath{\text{Ge\kern -0.1em V}}}
\newcommand{\pt}{\ensuremath{p_{\text{T}}}}
\newcommand{\met}{\ensuremath{E_{\text{T}}^{\text{miss}}}}

\makeatletter
\g@addto@macro\bfseries\boldmath
\makeatother

\title{From the Bottom to the Top -- Reconstruction of $\ttbar$ Events with Deep Learning}

\author{Johannes Erdmann, Tim Kallage, Kevin Kr\"oninger, Olaf Nackenhorst}

\affiliation{Lehrstuhl f\"ur Experimentelle Physik IV, TU Dortmund, Otto-Hahn-Stra{\ss}e 4a, 44227 Dortmund, Germany}

\emailAdd{johannes.erdmann@tu-dortmund.de}

\abstract{The reconstruction of top-quark pair-production ($t\bar{t}$) events is a prerequisite for many top-quark measurements. We use a deep neural network, trained with Monte-Carlo simulated events, to reconstruct $t\bar{t}$ decays in the lepton+jets final state. Comparing our approach to a widely-used kinematic fit, we find significant improvements in the correct assignment of jets to the partons from the decay, and we study the reconstruction performance of several kinematic top-quark properties. We document our workflow for the optimisation of the hyperparameters of the deep neural network. This workflow can be followed by experimental collaborations to retrain the network taking into account their detailed detector simulations.}


\arxivnumber{1907.11181}

\begin{document}
\maketitle
\flushbottom

\section{Introduction}
\label{sec:introduction}
At the Large Hadron Collider (LHC), top quarks have been produced copiously in Runs~1 and~2 and will be produced in even higher numbers in Run~3 and at the HL-LHC. The dominant process is the production of a top quark and an antitop quark (\ttbar) with a cross section of $832^{+40}_{-46}~\mathrm{pb}$~\cite{Czakon:2011xx} at a centre-of-mass energy of $\sqrt{s} = 13~\TeV$. Datasets comprised of millions of \ttbar\ events allow for measurements with unprecedented precision. Examples are measurements of Standard Model (SM) parameters, such as the top-quark mass~\cite{Aaboud:2018zbu,Sirunyan:2018gqx}, and measurements of differential cross sections~\cite{Aaboud:2017fha,Sirunyan:2019zvx}, which may unveil deviations from the SM predictions and provide evidence for physics beyond the SM.

The decay $\ttbar\rightarrow W^+bW^-\bar{b}$ has a branching ratio (BR) close to unity, so that this decay is almost exclusively used for top-quark measurements. Many measurements have been made in the lepton+jets decay channel, where one of the $W$ bosons decays leptonically and the other one decays hadronically. It is often chosen because of its significant BR ($\approx 30\%$ if electrons and muons are considered) and its relatively low background. Several important measurements in the lepton+jets channel require a reconstruction of the \ttbar\ decay, such as measurements of the top-quark mass~\cite{Aaboud:2018zbu,Sirunyan:2018gqx}, the $W$-helicity in top-quark decays~\cite{Aaboud:2016hsq,Khachatryan:2016fky}, the \ttbar\ charge asymmetry~\cite{Sirunyan:2017lvd}, and the top-quark width~\cite{Aaboud:2017uqq}.

The reconstruction of events in the lepton+jets channel, i.e.\ the assignment of jets to the four partons from the \ttbar\ decay, is often achieved with kinematic fitting. In various analyses, the \textsc{KLFitter} framework~\cite{Erdmann:2013rxa} has been used. In this algorithm, each possible jet-to-parton assignment is tested for consistency with the kinematics of the \ttbar\ decay chain using a maximum-likelihood approach. Wrong jet-to-parton assignments result in combinatorial background for the measurement and reduce its precision, so that a good \ttbar\ reconstruction is essential for such measurements.

Deep neural networks (DNNs) have been proposed for various applications in data analysis at the LHC~\cite{Guest:2018yhq}. One area in which they have shown to outperform algorithms that were previously used by the experimental collaborations is the identification of particles with a complex signature. Examples include the identification of jets from the hadronisation of $b$-quarks and $c$-quarks~\cite{ATL-PHYS-PUB-2017-013,Sirunyan:2017ezt}, the identification of high-momentum hadronically decaying top quarks~\cite{Aaboud:2018psm,CMS-DP-2017-049} and $W$, $Z$ and Higgs bosons~\cite{Aaboud:2018psm,Aaboud:2018wxv}, the classification of jets from the hadronisation of quarks and gluons~\cite{ATL-PHYS-PUB-2017-017,CMS-DP-2017-027}, and global particle identification~\cite{LHCb}.

In this paper, we study the reconstruction of \ttbar\ events with a DNN, and we compare its performance to \textsc{KLFitter} as an established benchmark algorithm. We use Monte-Carlo (MC) simulated events that are passed through a simplified simulation of a multipurpose detector at the LHC. We then train the DNN with basic kinematic information of the reconstructed objects and flavour-tagging information for the jets. For data analysis at an LHC experiment, the DNN would likely be retrained using the detailed detector simulations that are used at these experiments. We hence present a detailed workflow for retraining the DNN, focusing on the optimisation of the choice of hyperparameters. While automated algorithms for such optimisations exist~\cite{autohyp4, autohyp5, autohyp6, autohyp7}, the high dimensionality of the hyperparameter space and the large number of events needed result in a training that is computationally very intensive and hence often not practical in data analysis at LHC experiments. We present a simple sequential grid search, which in particular is parallelisable and helps to understand how the performance depends on certain hyperparameters. The proposed strategy is targeted at LHC experiments, but it may also be used for \ttbar\ reconstruction at future hadron- and lepton-collider experiments.

For the reconstruction of \ttbar\ events in association with a Higgs boson that decays via $H\rightarrow b\bar{b}$, a DNN was already shown to outperform a boosted decision tree as an alternative machine-learning approach~\cite{Erdmann:2017hra}. The added value of the work documented in this paper is threefold: We quantify the performance in reconstructing \ttbar\ events with a DNN, which has a wide range of applications, as outlined above. We compare the performance to the standard approach (kinematic fitting), which has been used for many analyses at the LHC~\cite{Aaboud:2018zbu,Sirunyan:2017lvd,Aaboud:2017uqq,Aaboud:2016hsq,Aaboud:2016bmk,Aad:2015nba,Chatrchyan:2016mqq,Khachatryan:2016fky,Sirunyan:2018gqx,Chatrchyan:2012cz}. We consider the simple sequential optimisation procedure that we perform and document ready-to-use for data analysis at LHC experiments.

This paper is structured as follows: In Section~\ref{sec:samples_selection}, we present the MC simulation samples and the event selection that we used. The training and optimisation of the DNN is the topic of Section~\ref{sec:optimisation}. In Section~\ref{sec:results}, we discuss the performance of the reconstruction using the DNN and compare it to the performance achieved with \textsc{KLFitter}. We present our conclusions in Section~\ref{sec:conclusions}.

\section{Monte Carlo samples and event selection}
\label{sec:samples_selection}
Samples of simulated events are produced for \ttbar\ production in proton--proton collisions at $\sqrt{s} = 13~\TeV$ with up to one additional parton in the final state using \textsc{MadGraph5\_aMC@NLO}~\cite{Alwall:2011uj} at leading order in the strong coupling constant and with a top-quark mass of 173~\GeV. \textsc{Pythia}~8~\cite{Sjostrand:2014zea} is used for the simulation of the top-quark decays to a $W$ boson and a $b$-quark, for the decay of the $W$ bosons, and for the simulation of the parton shower and the hadronisation. The $W$-boson decays are simulated so that one $W$ boson decays to an electron or muon and the corresponding neutrino, and the other $W$ boson decays to two quarks. Additional radiation in the matrix element is matched with the parton shower using the MLM procedure~\cite{Mangano:2006rw}. A simplified simulation of an LHC-type detector\footnote{We use the Delphes implementation of a CMS-like detector. However, the exact choice of the simplified detector simulation is believed to have little impact on the conclusions of this study.} is used, as implemented in \textsc{Delphes}~\cite{deFavereau:2013fsa}.

Electrons in the detector are required to have a transverse momentum, \pt, larger than 25~\GeV\ and an absolute value of the pseudorapidity, $\eta$, of at most 2.5. Muons in the detector are required to fulfil $\pt > 25~\GeV$ and $|\eta| < 2.4$. An isolation criterion is used for the charged leptons: the maximum energy around the electron or muon within a radius of $\Delta R = \sqrt{\left(\Delta \phi\right)^2 + \left(\Delta \eta\right)^2} = 0.5$ is required to be smaller than 12\% of the electron's \pt\ or 25\% of the muon's \pt, respectively. Calorimeter jets are reconstructed with the anti-$k_t$ algorithm~\cite{Cacciari:2008gp} with a radius parameter of 0.5 and required to fulfil $\pt > 25~\GeV$. Jets are $b$-tagged using a \pt-dependent efficiency for jets that are geometrically matched to a $b$-quark. The efficiency is $\approx 52\%$ at $\pt = 25~\GeV$, rises to $\approx 73\%$ at $\pt \approx 150~\GeV$, and decreases for higher \pt, such that at $\pt = 500~\GeV$ it is $\approx 55\%$. Mistagging efficiencies are applied to jets that are not geometrically matched to a $b$-quark, with higher mistagging efficiencies for jets that are matched to a $c$-quark than for jets that are matched to $u$-, $d$- or $s$-quarks or to gluons.

The lepton+jets final state is characterised by a charged lepton, missing transverse momentum from the undetected neutrino, and at least four jets, two of which are expected to originate from the hadronisation of a $b$-quark. In order to select events in this final state, we require exactly one reconstructed electron or muon in the events, and at least four jets with $|\eta| < 2.5$. We did not make requirements on the amount of \met\ and on the number of $b$-tagged jets. Out of 225\,467\,022 generated events, 36\,053\,266 events fulfil these selection criteria. As additional jets can be present in \ttbar\ events, for example due to initial-state or final-state radiation, we also study events with at least five and at least six jets. Out of all generated events, 19\,781\,437 events have at least five and 8\,654\,315 events have at least six jets.

The jets that are used for the reconstruction of the \ttbar\ decay chain are selected by the following procedure: In a first step, the $b$-tagged jets in the event are used. If more than two $b$-tagged jets are present, only the two $b$-tagged jets with the largest \pt\ are considered. In a second step, additional jets are considered in descending order in \pt, regardless of whether they are $b$-tagged or not. We consider either the first four, five or six jets that are selected using this procedure. We perform an unambiguous matching of the four partons from the \ttbar\ decay to the considered jets with the geometrical criterion $\Delta R\left(\mathrm{parton},\mathrm{jet}\right) < 0.3$. When considering the first four jets, the fraction of events in which all partons can be unambiguously matched to the considered jets is 19.6\%. When considering the first five or six jets in events with at least five or six jets, respectively, this fraction increases to 30.4\% and 38.3\%.

\section{Optimisation of the neural network}
\label{sec:optimisation}
For the training of the DNN, we only use events with an unambiguous matching of all partons to the considered jets, so that each matched jet is assigned the label of its matched parton. The goal of the \ttbar\ reconstruction algorithm is to predict the correct label for each of the considered jets, i.e.\ whether it corresponds to the $b$-quark from the hadronic top-quark decay, the $b$-quark from the leptonic top-quark decay, one of the light quarks from the hadronic $W$-boson decay, or whether it does not belong to the \ttbar\ decay chain and is unmatched. 

If all considered jets are assigned the correct label, the corresponding jet permutation is called ``correct permutation''. The prediction of the jet-to-parton assignment by an algorithm is called the ``predicted permutation''. The performance of a reconstruction algorithm can be quantified by the ``reconstruction efficiency'', defined as the fraction of unambiguously matched events in which the predicted permutation corresponds to the correct permutation. In unambiguously matched events with exactly four jets, there are $4! = 24$ permutations, but most reconstruction algorithms do not distinguish between the two light jets from the decay of the hadronic $W$ boson, so that only $4!/2 = 12$ independent permutations remain. In events with five jets, there are $5!/2 = 60$ independent permutations, and in events with six jets, there are $6!/2/2 = 180$ permutations, where the second factor of $1/2$ originates from the fact that also the two unmatched jets are not distinguished.

The DNN is trained as a binary classifier with the correct permutations as signal instances and with all other permutations as background instances\footnote{We also studied classifiers, for which only a fraction of the wrong permutations was included as background, which reduces the imbalance in the number of signal and background instances in the training. However, we found that this has a negative effect on the classifier performance and we concluded that it is beneficial to include all wrong permutations of an event during the training.}. This means that we use each event as often as permutations exist for that event. We sort the jets such that in the correct permutation, the first jet corresponds to the $b$-quark from the hadronically decaying top-quark, the second jet corresponds to the $b$-quark from the leptonically decaying top-quark, and the third and fourth jets correspond to the two jets from the hadronic $W$-boson decay. For the wrong permutations, the ordering is hence different. The input features for the DNN are the four-momenta of the considered jets, the information whether a jet is $b$-tagged or not, the three-momentum of the charged lepton (assuming that the mass of the charged lepton is negligible), and the magnitude and the azimuth, $\phi$, of the missing transverse momentum. As the azimuth is $2\pi$-continuous, for each object, $\sin\phi$ and $\cos\phi$ are used instead of $\phi$ itself. All variables are scaled using \textsc{scikit-learn}'s~\cite{Pedregosa:2012toh} StandardScaler, which transforms each variable so that it has a mean of zero and a standard deviation of unity. We treat events with electrons and events with muons together.

We use a fully-connected feedforward DNN implemented with \textsc{Keras}~\cite{chollet2015keras} using \textsc{TensorFlow}~\cite{tensorflow2015-whitepaper} as the back end. We optimise the hyperparameters of the DNN according to the workflow detailed below, where we use 60\% of the selected events for training (training sample), 20\% for the evaluation of the performance during the hyperparameter optimisation (validation sample) and 20\% for the final evaluation of the performance (test sample). For the minimisation during the training, we use the Adam optimiser~\cite{Kingma:2014vow} with binary cross entropy as loss function. The rectified linear unit is used as the activation function in the hidden layers, and the sigmoid function is used as activation function in the output layer. The batch size and the learning rate are hyperparameters of the Adam optimiser that define the sample size over which gradients are calculated during stochastic gradient descent and the step size in the optimisation, respectively. We choose the batch size to be a multiple of the number of permutations in order to always include all permutations of a single event in the batch\footnote{We ensure that all permutations that belong to one event are always included in the same batch. We found that the classifier performance decreases if they are randomly distributed over batches.}. In each case we train for a maximum number of 200 iterations on the training sample (epochs). The measure that we use to evaluate the performance of a trained DNN is the reconstruction efficiency. The best network of one training is chosen using the epoch with the largest reconstruction efficiency on the validation sample. If the value of the loss function on the validation sample does not improve over 20 epochs, the training is stopped in order to avoid strong overfitting. We further prevent overfitting by regularising the DNN using dropout~\cite{dropout} and the L$_2$ norm. Regularisation with dropout is based on randomly removing a certain fraction (called ``dropout rate'') of nodes in the hidden layers during the training. Because using dropout requires wider network structures and we keep the dropout rate constant in the hidden layers, we increase the number of nodes in the hidden layers to compensate for the number of dropped nodes in order to avoid instabilities during the training, which can otherwise occur in particular in the last and more narrow layers.
In the L$_2$ regulariser, a penalty term is added to the loss function that is given by the squared sum of all weights in a hidden layer, multiplied by a strength factor (the L$_2$ parameter). We divide the value of the L$_2$ parameter in a given layer by two with respect to the previous hidden layer, which effectively reduces the strength of the L$_2$ regularisation from the first to the last hidden layer.

The workflow for the optimisation of the hyperparameters follows a sequence of two-dimensional grid searches:
\begin{enumerate}
\item Optimisation of the DNN structure: We vary the number of hidden layers and the number of nodes in the first hidden layer. The number of nodes in the following hidden layers are always half of the number of the nodes in the previous layer. The batch size and the value of the learning rate are kept constant and are chosen manually from a few trial trainings as 6000 and $10^{-3}$, respectively. Based on the results of all trainings, we choose a network structure with a good performance and a moderate amount of overfitting, which indicates that the structure has a capacity that is large enough for our task.
\item Optimisation of the hyperparameters of the optimiser: For the chosen structure from step 1, we vary the batch size and the learning rate. Based on the results of all trainings, we choose the hyperparameters that result in the best performance.
\item Optimisation of the regulariser: Since we do not apply regularisation in steps 1 and 2, the DNN may overfit to the training sample. We use the hyperparameters from the previous optimisation steps and we additionally optimise the dropout fraction and the L$_2$ parameter to choose a final set of hyperparameters that balances performance and overfitting.
\end{enumerate}

We optimise DNNs separately for events with at least four jets, events with at least five jets and events with at least six jets. With an increasing number of jets, the classification problem becomes more and more challenging due to the increasing number of possible jet permutations. We discuss the steps of the optimisation procedure for events with at least four jets in detail in the following. The results of the optimisation of the DNNs for at least five and at least six jets are presented afterwards. When we train the DNNs with events with at least four, five or six jets, we always consider the corresponding number of jets for the jet permutations (four, five or six), following the ordering described at the end of Section~\ref{sec:samples_selection}.

\begin{figure}
  \centering
  \includegraphics[width=0.65\textwidth]{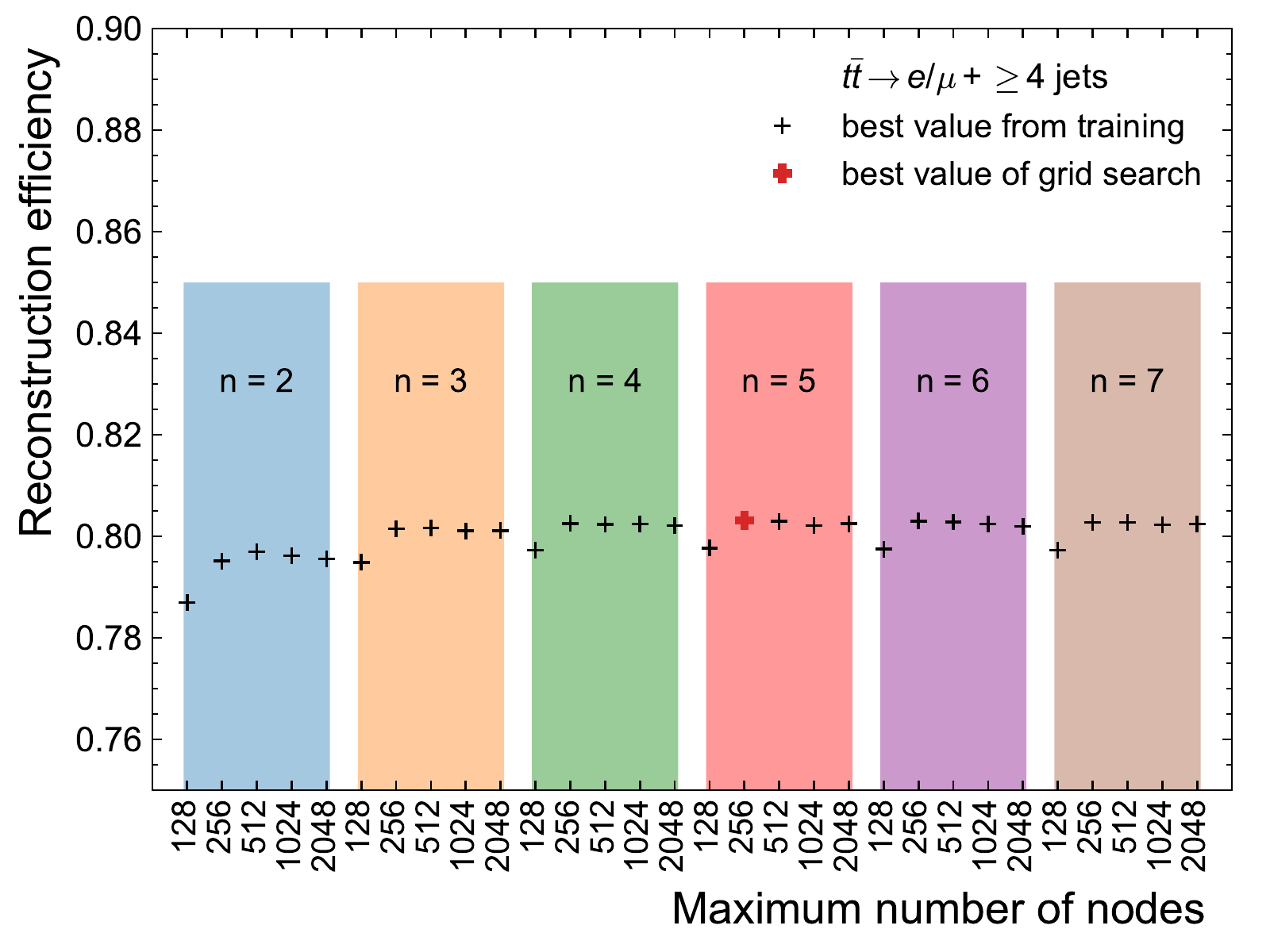}
  \caption{Reconstruction efficiency for events with at least four jets for different choices of the number of hidden layers ($n$) and the number of nodes in the first hidden layer (``maximum number of nodes''). The set of hyperparameters that results in the best value of the reconstruction efficiency is highlighted with a bold marker.}
  \label{fig:bestreco_4j_1}
\end{figure}

\begin{figure}[p]
  \centering
  \subfloat[]{\includegraphics[width=0.65\textwidth]{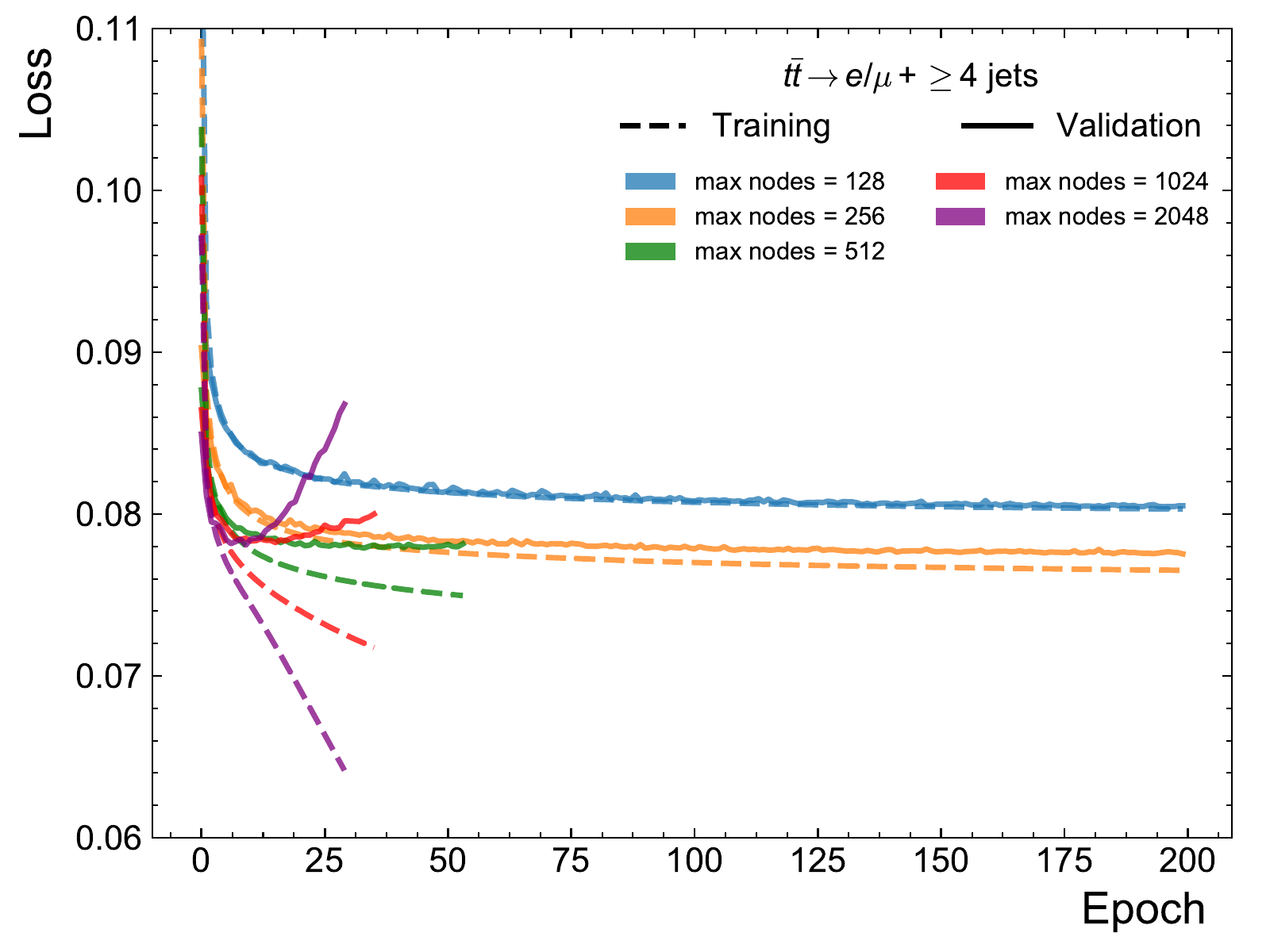}}\\
  \subfloat[]{\includegraphics[width=0.65\textwidth]{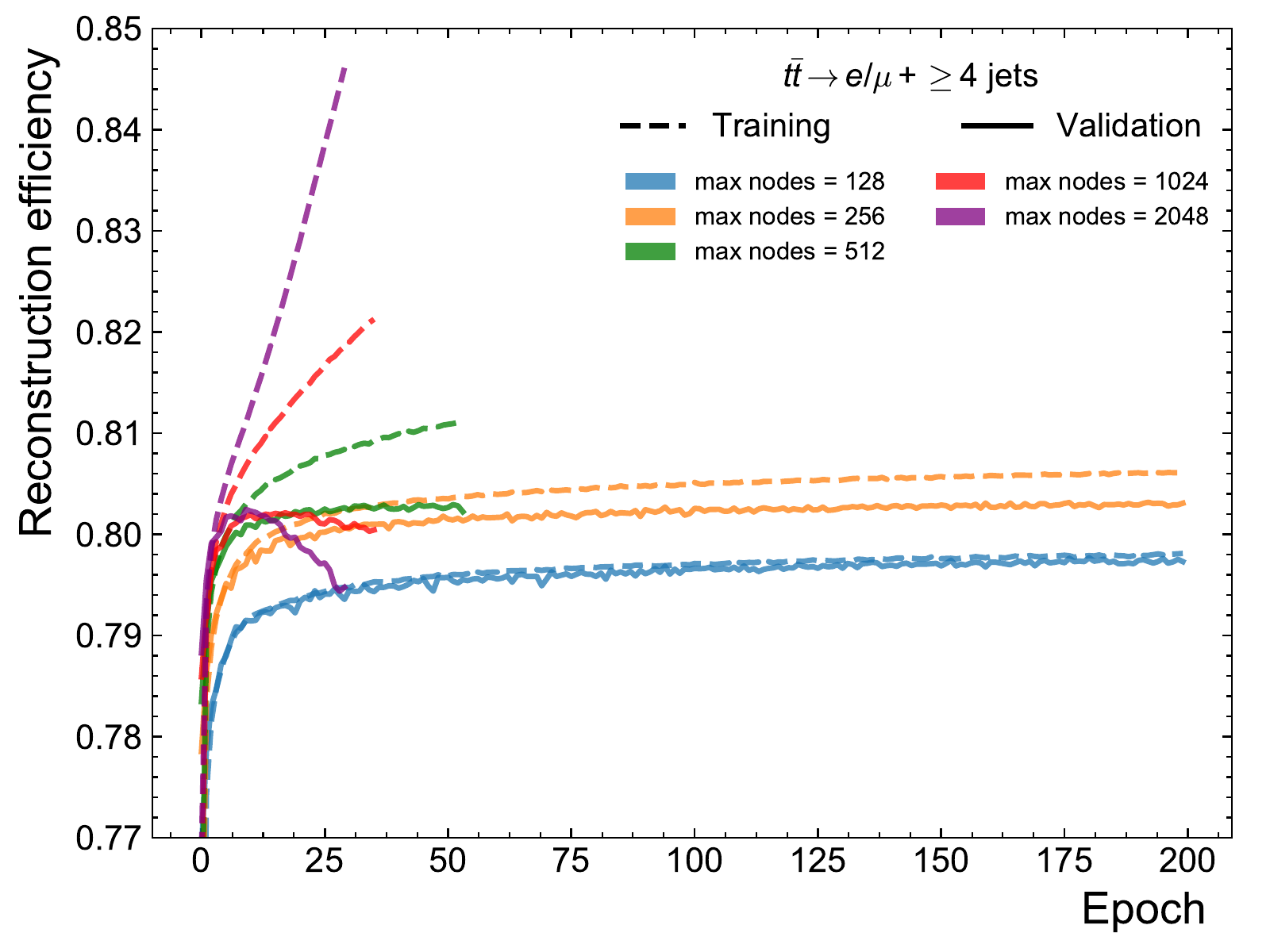}}
  \caption{Value of (a) the loss function and (b) the reconstruction efficiency as a function of the training epoch for events with at least four jets for the network with 5 hidden layers and different choices of the number of nodes in the first hidden layer (``max. nodes''). The learning rate is set to a value of $10^{-3}$ and the batch size to a value of 6000. The values that are calculated with the training sample are shown as dashed lines and the values calculated with the validation sample are shown as solid lines.}
\label{fig:metric_4j_1}
\end{figure}

1. Optimisation of the DNN structure: In Figure~\ref{fig:bestreco_4j_1}, the reconstruction efficiency is shown for networks where the number of hidden layers is varied from 2 to 7 in steps of one and the number of nodes in the first hidden layer is varied using the values 128, 256, 512, 1024 and 2048. The reconstruction efficiency for networks with only 2 hidden layers and for networks with only 128 nodes in the first hidden layer is lower than the reconstruction efficiency of networks with a larger number of hidden layers or number of nodes in the first hidden layer. We conclude that the capacity of such small networks is not large enough for the classification task. The best reconstruction efficiency is seen for the network with 5 hidden layers and 256 nodes in the first layer. The values of the loss function and the reconstruction efficiency are shown in Figures~\ref{fig:metric_4j_1}(a) and~(b) as a function of the training epoch for networks with 5 hidden layers and different choices of the number of nodes in the first hidden layer. The network with the best reconstruction efficiency was trained for the full 200 epochs and shows only a slight tendency of overfitting. Networks with a larger number of nodes in the first layer, however, enter the regime of overfitting. For further optimisation, we choose a network structure with 5 hidden layers and double the number of nodes in the first hidden layer (512), which shows slight overfitting. We choose this network structure, because its larger capacity promises a better performance when the network is regularised in step 3.

\begin{figure}[p]
  \centering
  \includegraphics[width=0.65\textwidth]{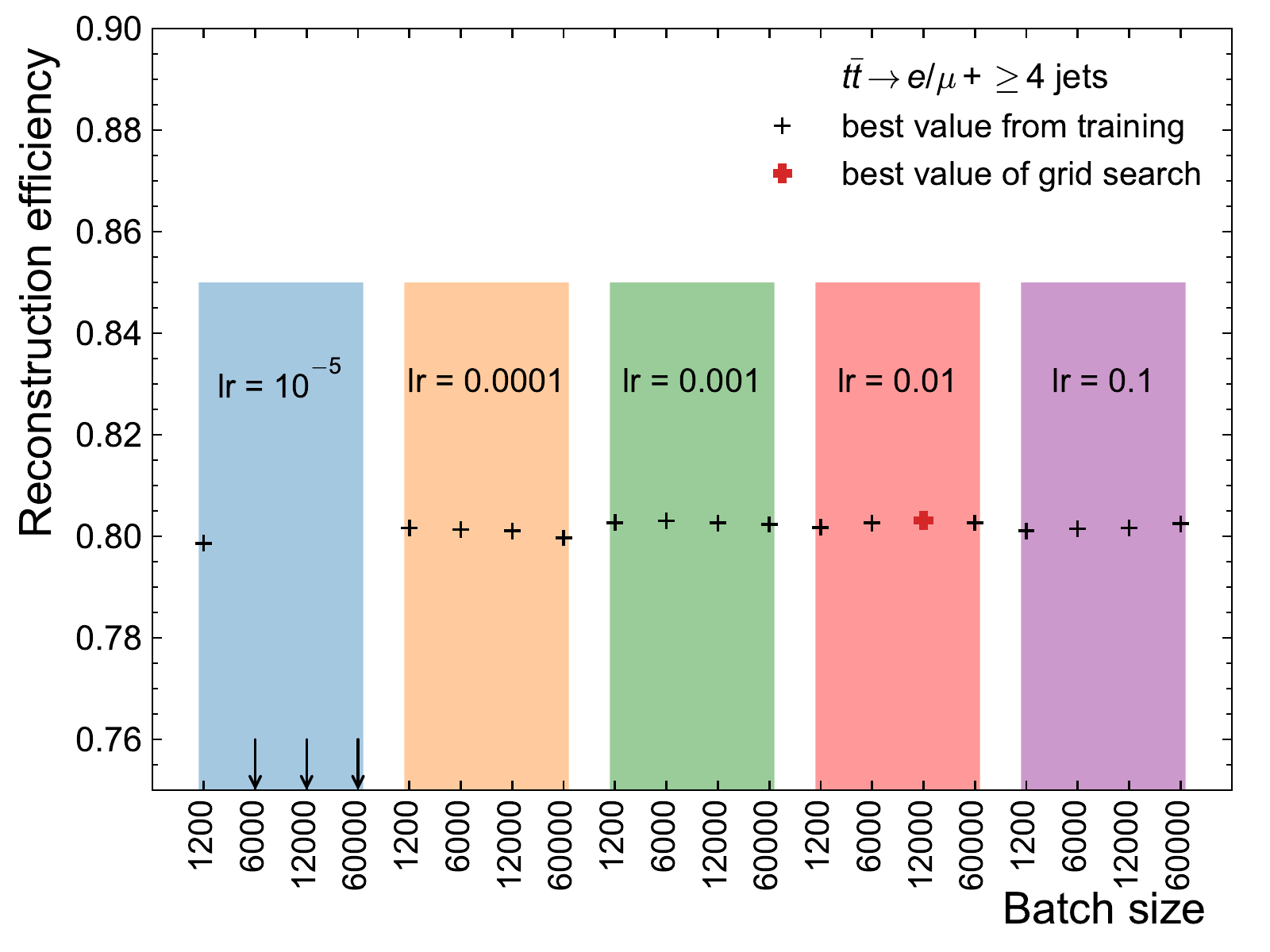}
  \caption{Reconstruction efficiency for events with at least four jets for different choices of the batch size and the learning rate (``lr''). The number of hidden layers is set to 5 and the number of nodes in the first hidden layer is set to 512. The set of hyperparameters that results in the best value of the reconstruction efficiency is highlighted with a bold marker. The downward-pointing arrows indicate that the reconstruction efficiency is lower than the scale of the $y$-axis.}
  \label{fig:bestreco_4j_2}
  \includegraphics[width=0.65\textwidth]{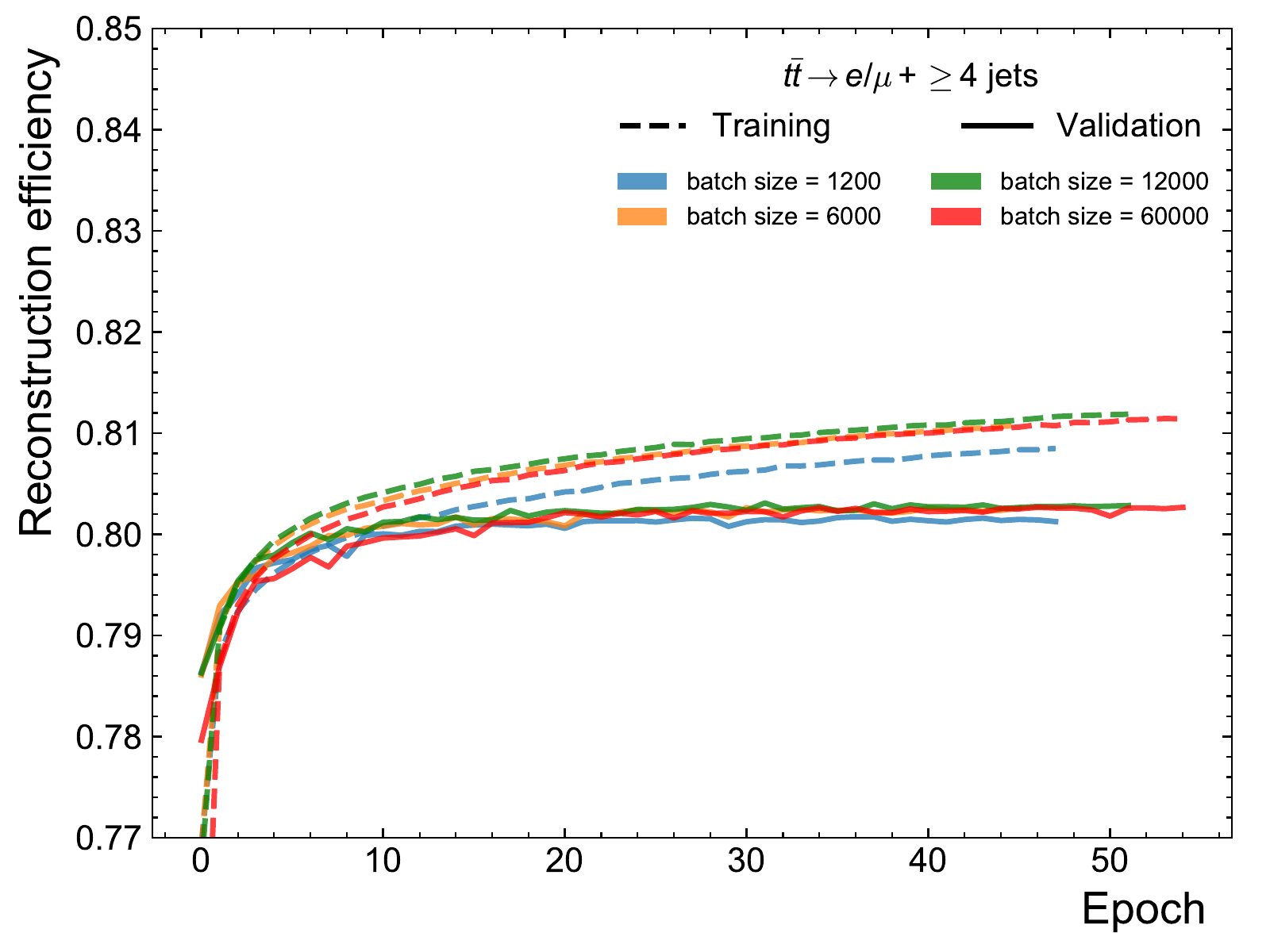}
  \caption{Value of the reconstruction efficiency as a function of the training epoch for events with at least four jets for the network with a learning rate of 0.01 and different values of the batch size. The values that are calculated with the training sample are shown as dashed lines and the values calculated with the validation sample are shown as solid lines.}
  \label{fig:metric_4j_2}
\end{figure}

2. Optimisation of the hyperparameters of the optimiser: In Figure~\ref{fig:bestreco_4j_2}, the reconstruction efficiency is shown for the network with 5 hidden layers and 512 nodes in the first hidden layer if the learning rate is varied using the values $10^{-5}$, 0.0001, 0.001 and 0.1 and the batch size is varied using the values 1200, 6000, 12\,000 and 60\,000. Except for the case of too large batch sizes for a learning rate of $10^{-5}$, the reconstruction efficiencies for all other combinations of learning rate and batch size are very similar. The DNN with a learning rate of 0.01 and a batch size of 12\,000 is chosen, as it shows the best performance. A relatively large learning rate in combination with a relatively large batch size has the additional advantage of a smaller training time compared to trainings with small learning rates and small batch sizes. The reconstruction efficiency as a function of the training epoch for a learning rate of 0.01 and for different values of the batch size is shown in Figure~\ref{fig:metric_4j_2}. All trained networks show some overfitting.

\begin{figure}
  \centering
  \includegraphics[width=0.65\textwidth]{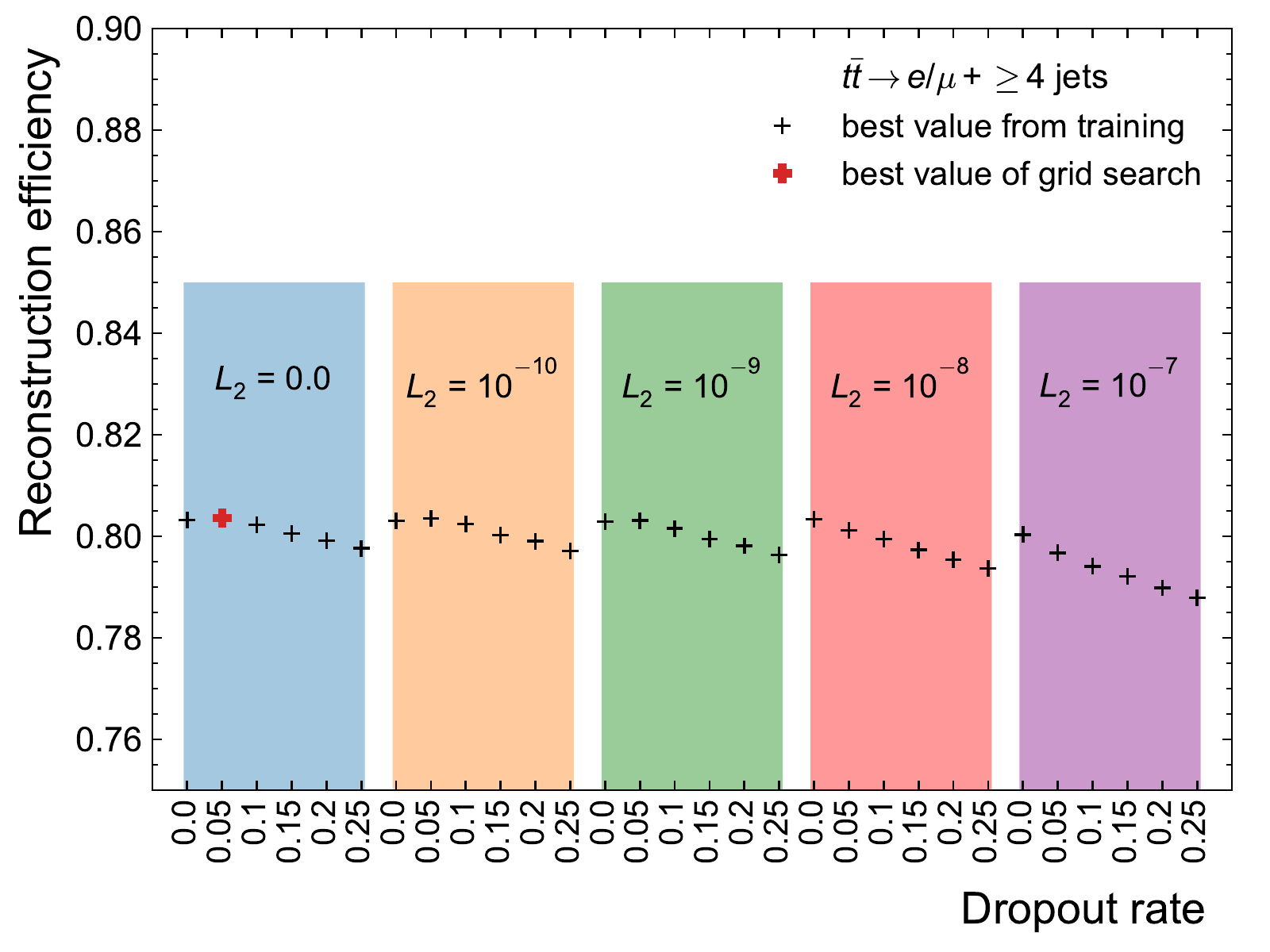}
  \caption{Reconstruction efficiency for events with at least four jets for different choices of the dropout rate (``dropout'') and the parameter of the L$_2$ regularisation (``l2''). The number of hidden layers is set to 5, the number of nodes in the first hidden layer to 512, the learning to 0.01 and the batch size to 12\,000. The set of hyperparameters that results in the best value of the reconstruction efficiency is highlighted with a bold marker.}
  \label{fig:bestreco_4j_3}
\end{figure}

\begin{figure}
  \centering
  \includegraphics[width=0.65\textwidth]{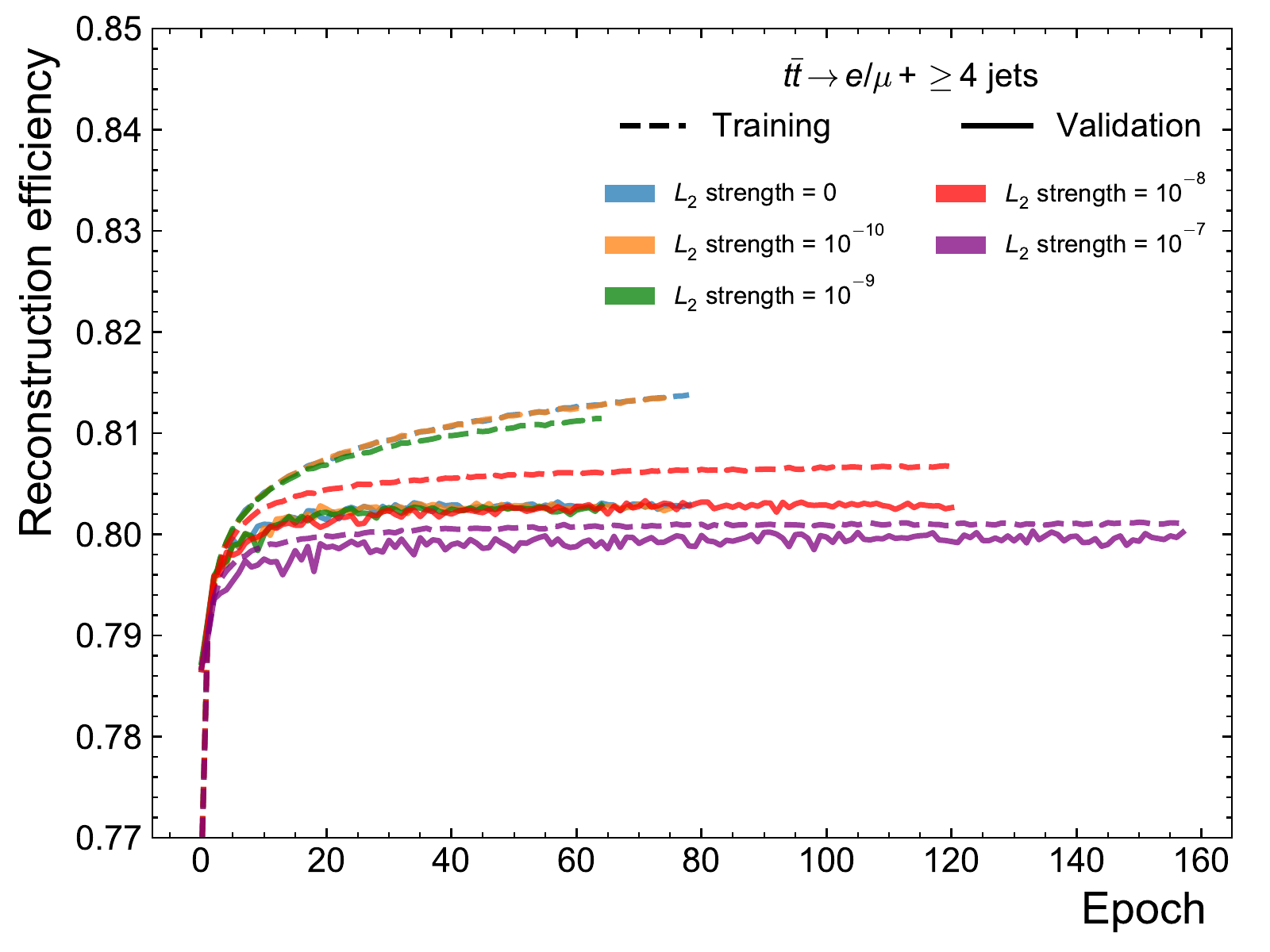}
  \caption{Value of the reconstruction efficiency as a function of the training epoch for events with at least four jets for the network that is regularised only with the L$_2$ regulariser for different values of the L$_2$ parameter. The values that are calculated with the training sample are shown as dashed lines and the values calculated with the validation sample are shown as solid lines.}
  \label{fig:metric_4j_3}
\end{figure}

3. Optimisation of the regulariser: In Figure~\ref{fig:bestreco_4j_3}, the reconstruction efficiency is shown for the network with 5 hidden layers and 512 nodes in the first hidden layer and with a learning rate of 0.01 and a batch size of 12\,000 if the dropout rate is varied from 0\% to 25\% in steps of 5\% and the L$_2$ parameter is varied using the values 0, 10$^{-10}$, 10$^{-9}$, 10$^{-8}$ and 10$^{-7}$. The reconstruction efficiency decreases if the network is too strongly regularised, i.e.\ the dropout rates and the values of the L$_2$ parameter are too large. Networks with a mild regularisation, however, show a similar performance as the network without regularisation. One such network with a mild regularisation uses L$_2$ regularisation with a parameter of $10^{-8}$ and no dropout regularisation (dropout rate of 0). We chose this regularisation with L$_2$ over the setup with the best reconstruction efficiency using dropout, because the difference in performance between these setups is marginal and the difference between the reconstruction efficiency on the training and the validation datasets was found to be smaller in this case. The reconstruction efficiency as a function of the training epoch for the networks that are only regularised using L$_2$ is shown in Figure~\ref{fig:metric_4j_3} for different values of the L$_2$ parameter. Compared to the performance of the DNN without regularisation (Figure~\ref{fig:metric_4j_2}), overtraining is reduced if an L$_2$ parameter of $10^{-8}$ is chosen and the reconstruction efficiency is slightly higher than without regularisation.

\begin{figure}
  \centering
  \includegraphics[width=0.65\textwidth]{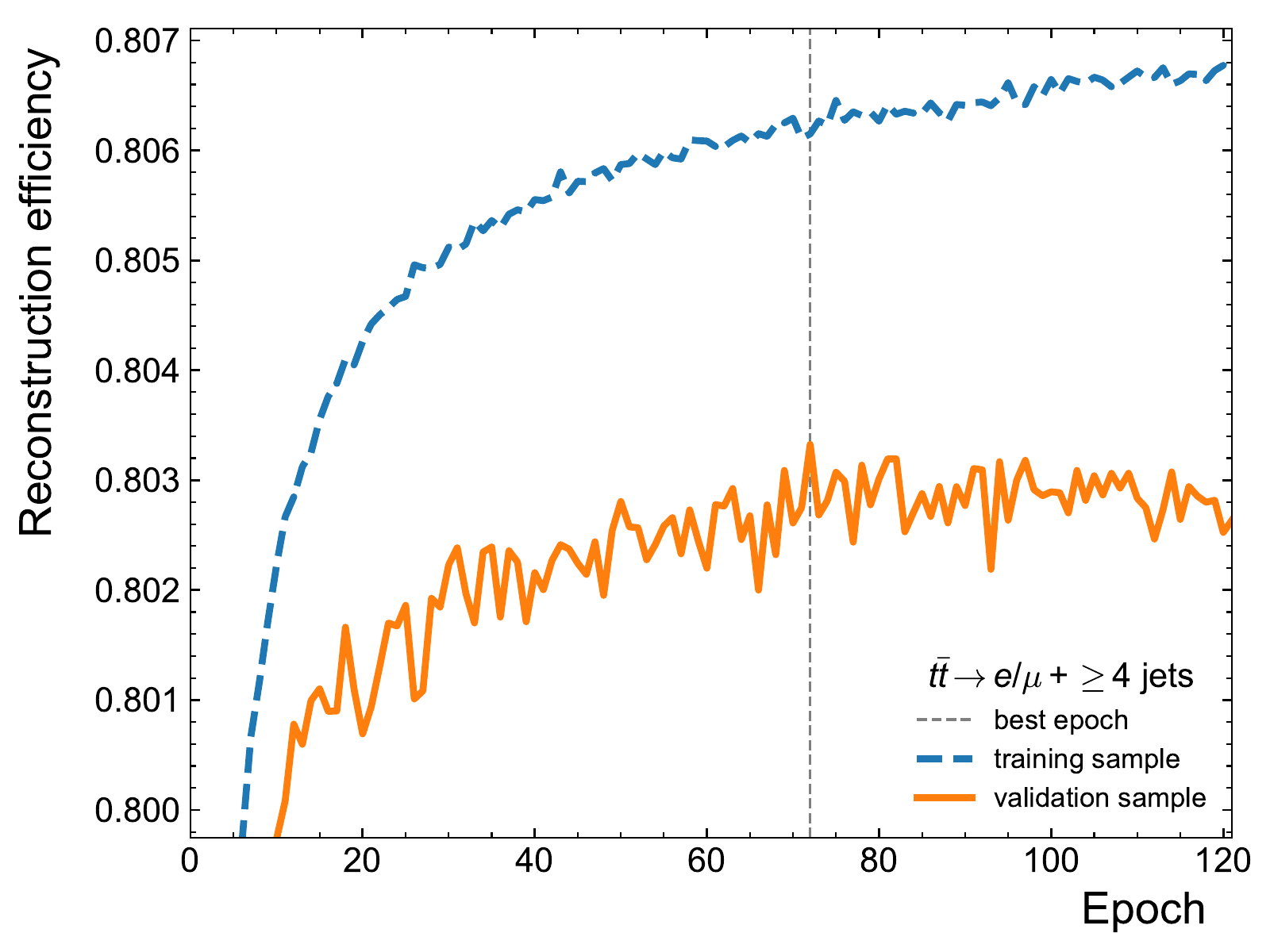}
  \caption{Value of the reconstruction efficiency as a function of the training epoch for events with at least four jets for the network with the final optimised hyperparameters. The values that are calculated with the training sample are shown as dashed line and the values calculated with the validation sample are shown as solid line. The best training is indicated by the vertical line.}
  \label{fig:bestreco_4j_4}
\end{figure}

The final optimised hyperparameters for the DNN trained with events with at least four jets are shown in Table~\ref{tab:4jets}. The reconstruction efficiency as a function of the training epoch is shown in Figure~\ref{fig:bestreco_4j_4}. The training with the best reconstruction efficiency on the validation sample is chosen, which is indicated by the vertical line. Overall, the reconstruction efficiency increases from 80.32\% to 80.36\% from step~1 to step~3. The increase is small due to a good initial guess of the hyperparameters, which is not guaranteed when the network would be retrained with a more detailed detector simulation.

\begin{figure}[p]
  \centering
  \includegraphics[width=0.65\textwidth]{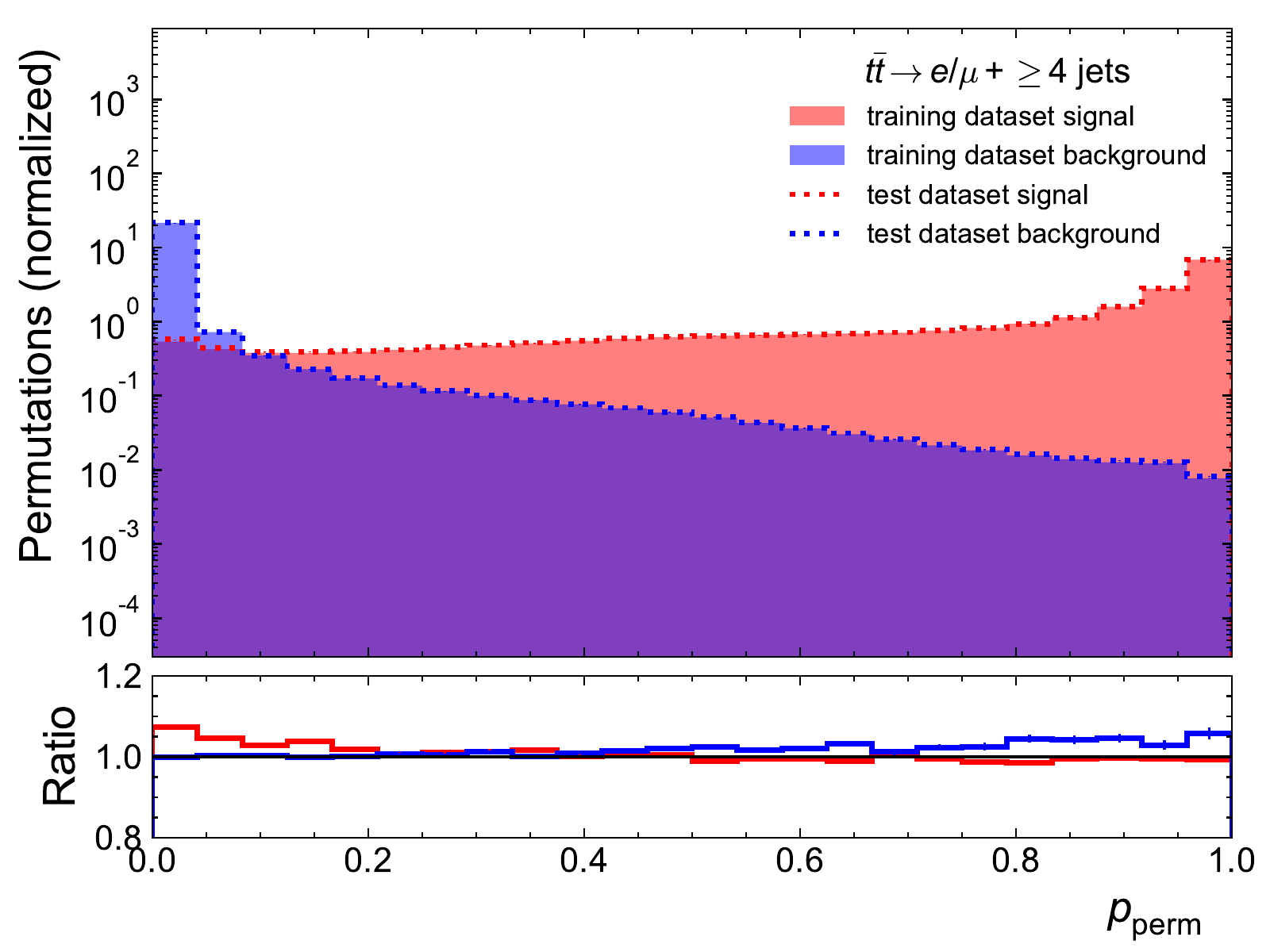}
  \caption{Distribution of the permutation probability ($p_{\mathrm{perm}}$) for events with at least four jets for the network with the final optimised hyperparameters. The distributions for the correct permutation (signal) and for all wrong permutations (background) are shown separately if evaluated on the training sample and on the test sample. The ratio of the distributions from training and test samples is also shown.}
  \label{fig:overtraining_4j}
\end{figure}

In order to test for residual overtraining in the optimised DNN output, the distribution of the ``permutation probability'' in the training sample and in the test sample is compared. The permutation probability is defined as the ratio of the DNN output for the predicted permutation divided by the sum of the DNN outputs of all permutations in an event. The distributions of the permutation probability in the training and the test sample are shown for the correct and for all wrong jet permutations in Figure~\ref{fig:overtraining_4j}. The comparison of training-sample and test-sample distributions shows that only a small amount of residual overtraining is present in the optimised DNN.

\begin{table}
\centering
\caption{Optimised hyperparameters for events with at least four jets.}
\begin{tabular}{lc}
\toprule
Hyperparameter & Value \\
\midrule
Hidden layers & 5\\
Number of nodes & \\
in the hidden layers & 512, 256, 128, 64, 32\\
Batch size & 12\,000\\
Learning rate & 0.01\\
L$_2$ parameter & $10^{-8}$\\
Dropout rate & 0\\
\bottomrule
\end{tabular}
\label{tab:4jets}
\end{table}

\begin{figure}[p]
  \centering
  \subfloat[]{\includegraphics[width=0.65\textwidth]{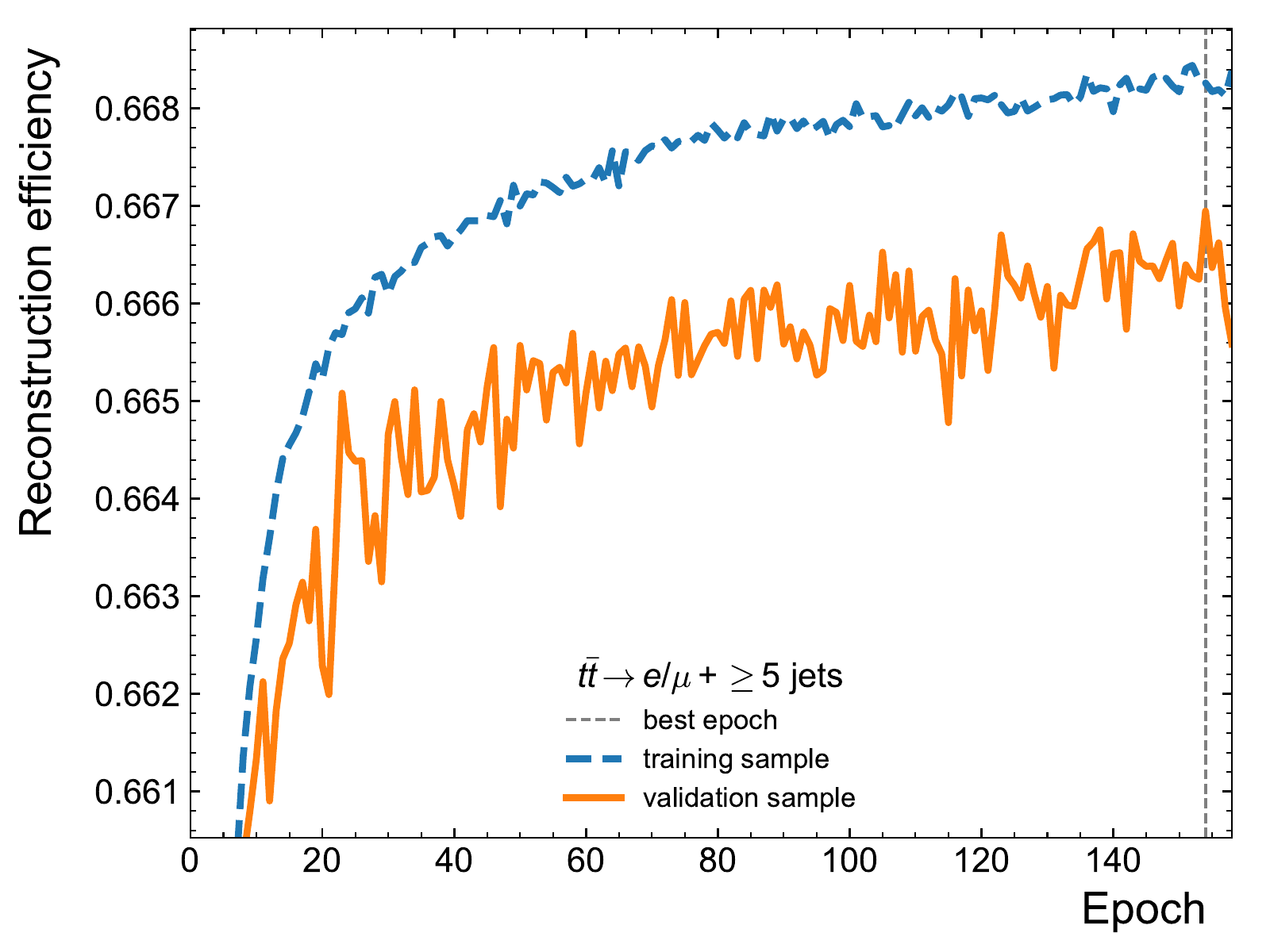}}\\
  \subfloat[]{\includegraphics[width=0.65\textwidth]{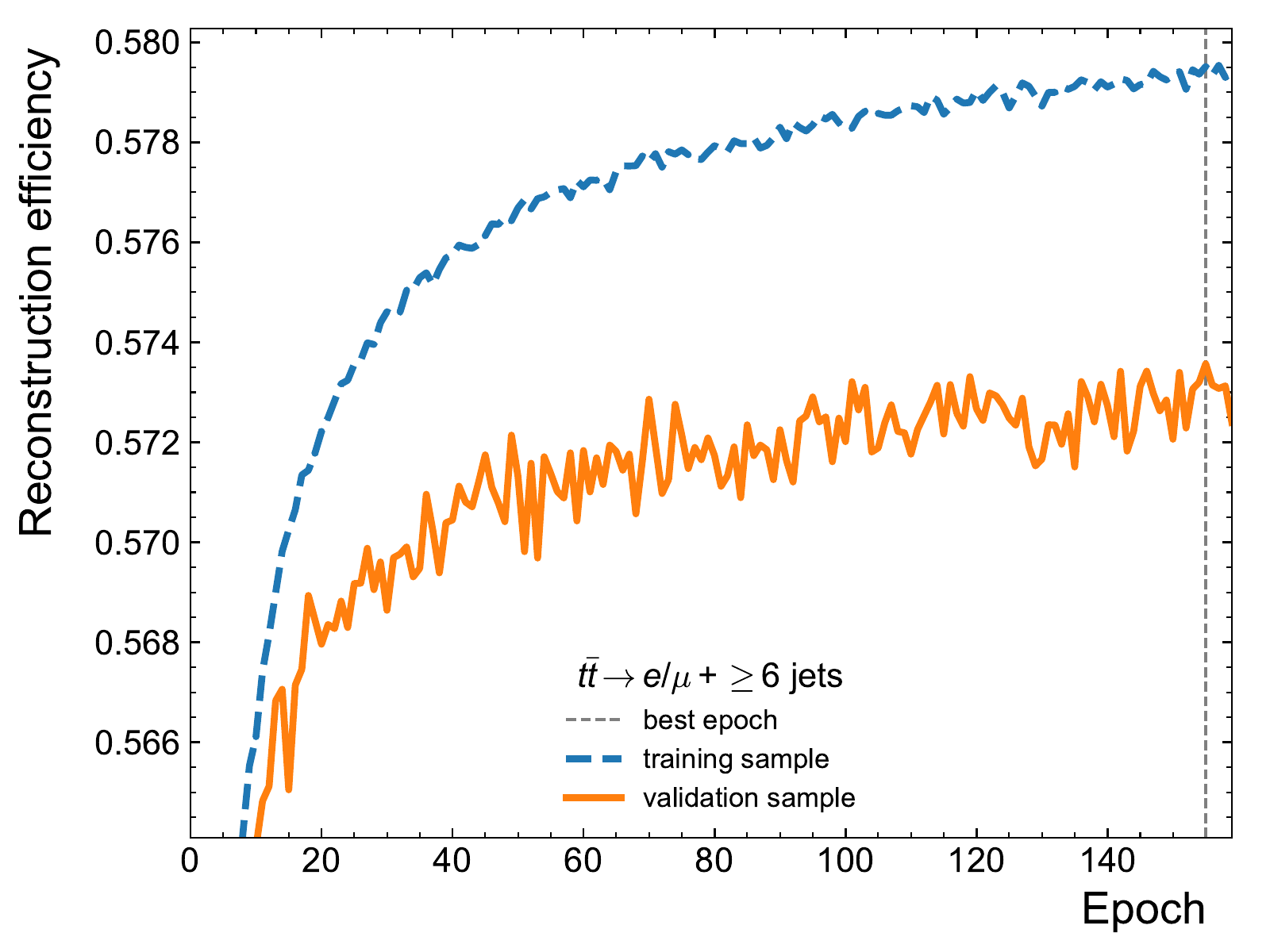}}
  \caption{Value of the reconstruction efficiency as a function of the training epoch for events (a) with at least five jets and (b) with at least six jets for the network with the final optimised hyperparameters. The values that are calculated with the training sample are shown as dashed line and the values calculated with the validation sample are shown as solid line. The best trainings are indicated by the vertical line.}
  \label{fig:bestreco_56j}
\end{figure}

\begin{figure}[p]
  \centering
  \subfloat[]{\includegraphics[width=0.65\textwidth]{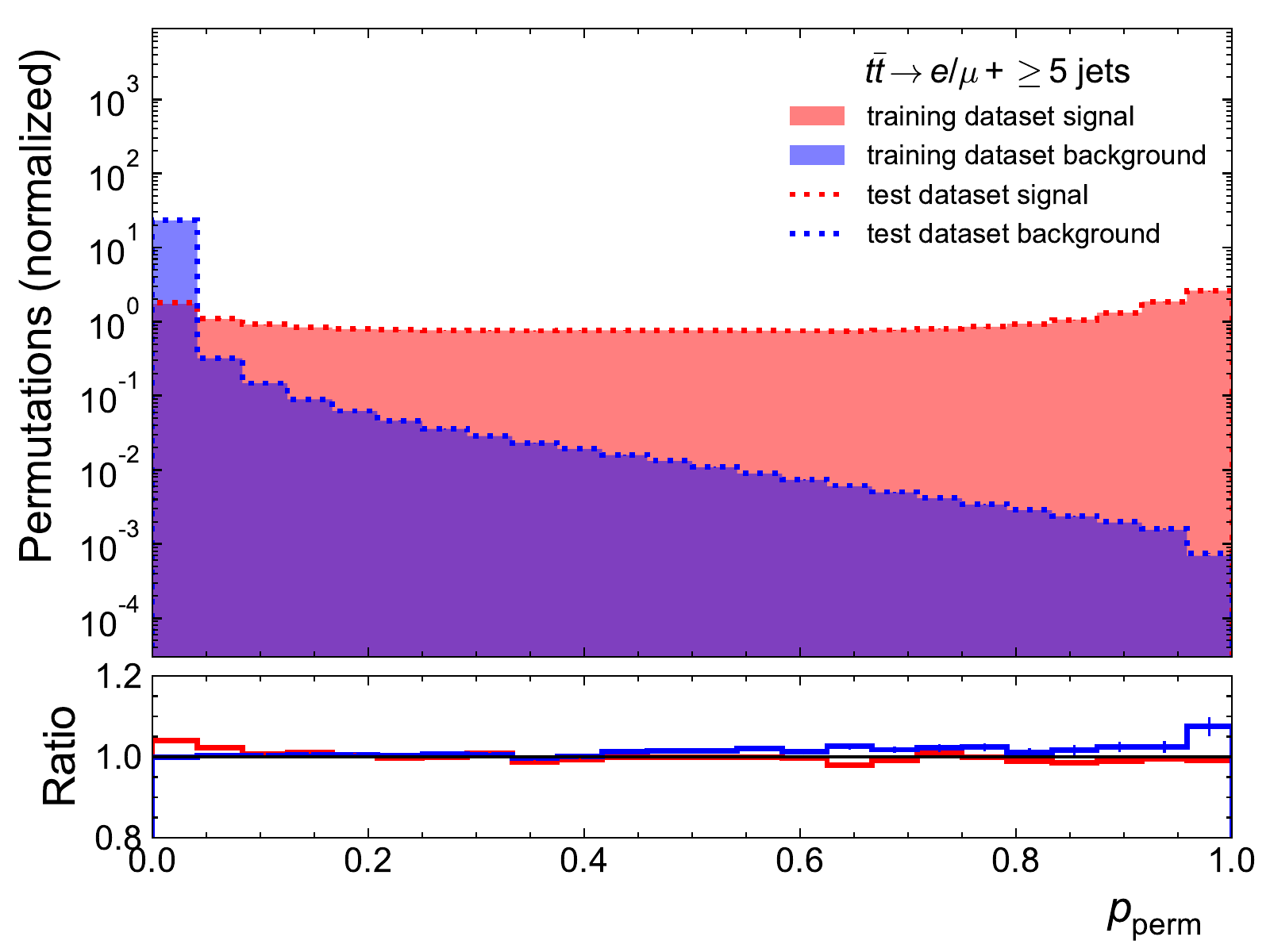}}\\
  \subfloat[]{\includegraphics[width=0.65\textwidth]{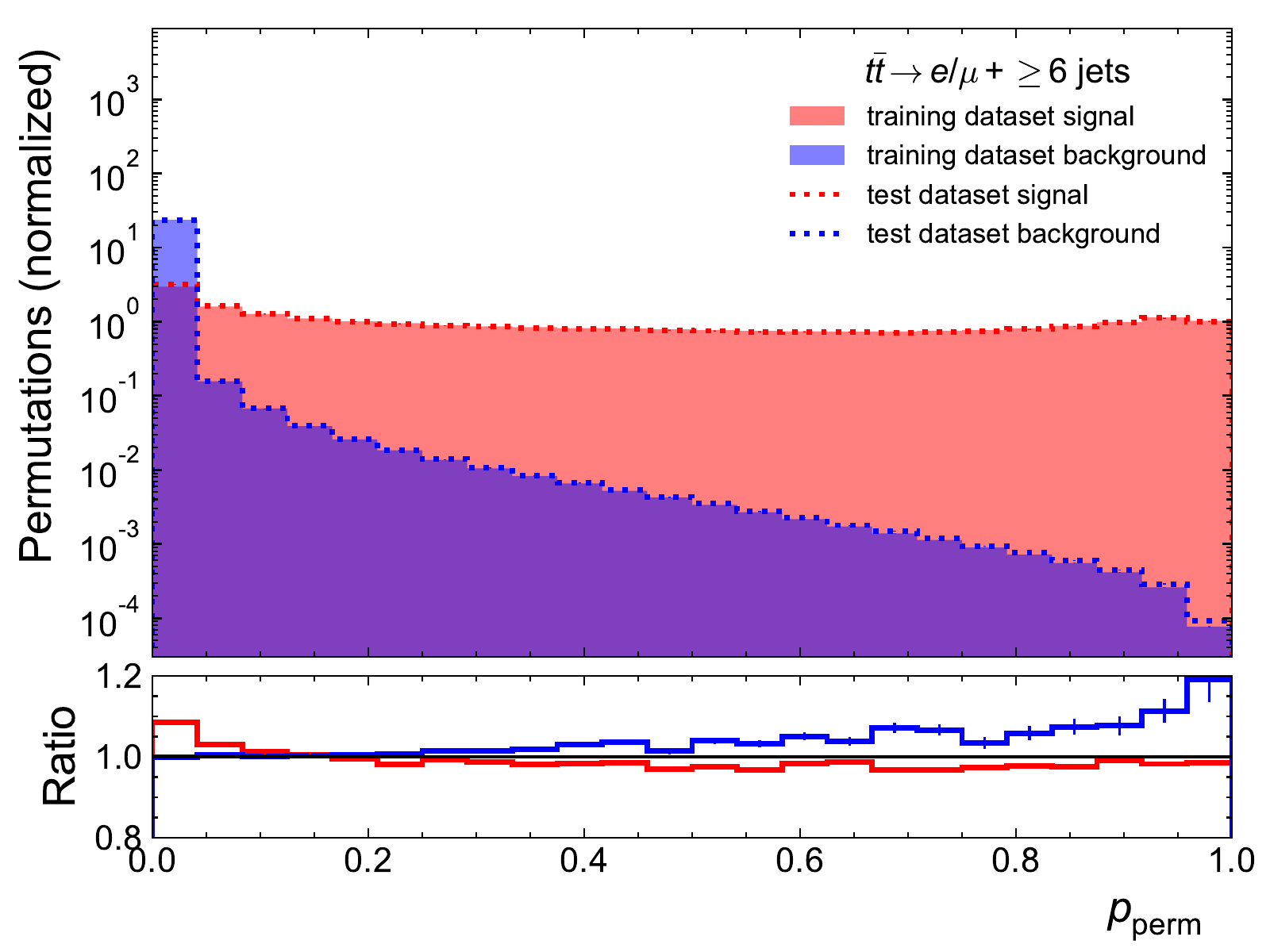}}
  \caption{Distribution of the permutation probability ($p_{\mathrm{perm}}$) for events (a) with at least five jets and (b) with at least six jets for the network with the final optimised hyperparameters. The distributions for the correct permutation (signal) and for all wrong permutations (background) are shown separately if evaluated on the training sample and on the test sample. The ratio of the distributions from training and test samples is also shown.}
  \label{fig:overtraining_56j}
\end{figure}

For the optimisation of the DNNs for events with at least five jets and at least six jets, we follow the same procedure as discussed above. For the optimisation of the regulariser, however, we only use the L$_2$ regulariser, as this is found to be sufficient for regularising the four-jet DNN discussed above. The hyperparameters of the final optimised networks are shown in Tables~\ref{tab:5jets} and~\ref{tab:6jets}. In both cases, similar values are found for the number of hidden layers, the number of nodes in the first hidden layer, the learning rate and the value of the L$_2$ parameter, as in the four-jet case. The optimal batch size, however, is found to be larger, which is expected because the number of permutations (and hence the number of wrong permutations) per event is larger in the five- and six-jet cases. The reconstruction efficiencies as a function of the training epoch are shown in Figure~\ref{fig:bestreco_56j}. The trainings with the best reconstruction efficiency on the validation sample are chosen (indicated by the vertical line in the figures). The distributions of the permutation probability on the training and on the test sample for the correct and for all wrong jet permutations are shown in Figure~\ref{fig:overtraining_56j}. The comparison of training-sample and test-sample distributions shows that only a small amount of residual overtraining is present in the optimised DNN for the five-jet case. In the six-jet case, the overtraining is larger than in the other two cases, since the network is less regularised.

\begin{table}
\centering
\caption{Optimised hyperparameters for events with at least five jets.}
\begin{tabular}{lc}
\toprule
Hyperparameter & Value \\
\midrule
Hidden layers & 6\\
Number of nodes & \\
in the hidden layers & 512, 256, 128, 64, 32, 16\\
Batch size & 60\,000\\
Learning rate & 0.01\\
L$_2$ parameter & $10^{-8}$\\
\bottomrule
\end{tabular}
\label{tab:5jets}
\end{table}

\begin{table}
\centering
\caption{Optimised hyperparameters for events with at least six jets.}
\begin{tabular}{lc}
\toprule
Hyperparameter & Value \\
\midrule
Hidden layers & 5\\
Number of nodes & \\
in the hidden layers & 512, 256, 128, 64, 32\\
Batch size & 60\,000\\
Learning rate & 0.01\\
L$_2$ parameter & $10^{-9}$\\
\bottomrule
\end{tabular}
\label{tab:6jets}
\end{table}

\clearpage

\section{Results}
\label{sec:results}
We evaluate the performance of the optimised DNN and we compare it to the performance obtained with the \textsc{KLFitter} algorithm. The \textsc{KLFitter} algorithm uses a maximum-likelihood approach based on Breit-Wigner distributions of the top-quark and $W$-boson invariant masses that are convolved with transfer functions for the jet and lepton energies, as well as for the $x$- and $y$-components of the missing transverse momentum. The transfer functions model the detector response and resolution of these quantities. In each jet permutation, a likelihood is maximised by varying the free parameters in a kinematic fit\footnote{In the case of Gaussian approximations for the Breit-Wigner distributions and the transfer functions, this approach reduces to a traditional kinematic fit using a $\chi^2$ function.}, which are: the energies of the four partons from the \ttbar\ decay, the energy of the lepton from the leptonic $W$-boson decay, and the $x$-, $y$- and $z$-component of the neutrino four-momentum. For a competitive comparison, we have chosen a \textsc{KLFitter} configuration that results in a large probability for the correct prediction of the jet-to-parton assignment: The central value of the Breit-Wigner distribution for the top-quark invariant mass is fixed to the value that is used in the generation of the MC samples (``fixed top-quark mass''), and $b$-tagging information for the jets is used by vetoing permutations in which a $b$-tagged jet is in a position of a light quark\footnote{As for events with more than two $b$-tagged jets, no permutation would survive this veto, in such cases only permutations in which a non-$b$-tagged jet is in the position of a $b$-quark are vetoed.}. For an additional comparison, we also show the performance of a \textsc{KLFitter} configuration where the top-quark invariant mass is not fixed but the reconstructed masses of both top quarks are required to be compatible with each other (``floating top-quark mass''). Depending on the number of $b$-tagged jets, the reconstruction with the \textsc{KLFitter} algorithm can result in less independent jet permutations as obtained for the reconstruction with the DNN. The jet permutation with the largest value of the likelihood is the \textsc{KLFitter} prediction for the jet-to-parton assignment.

\begin{figure}[p]
  \centering
  \subfloat[]{\includegraphics[width=0.49\textwidth]{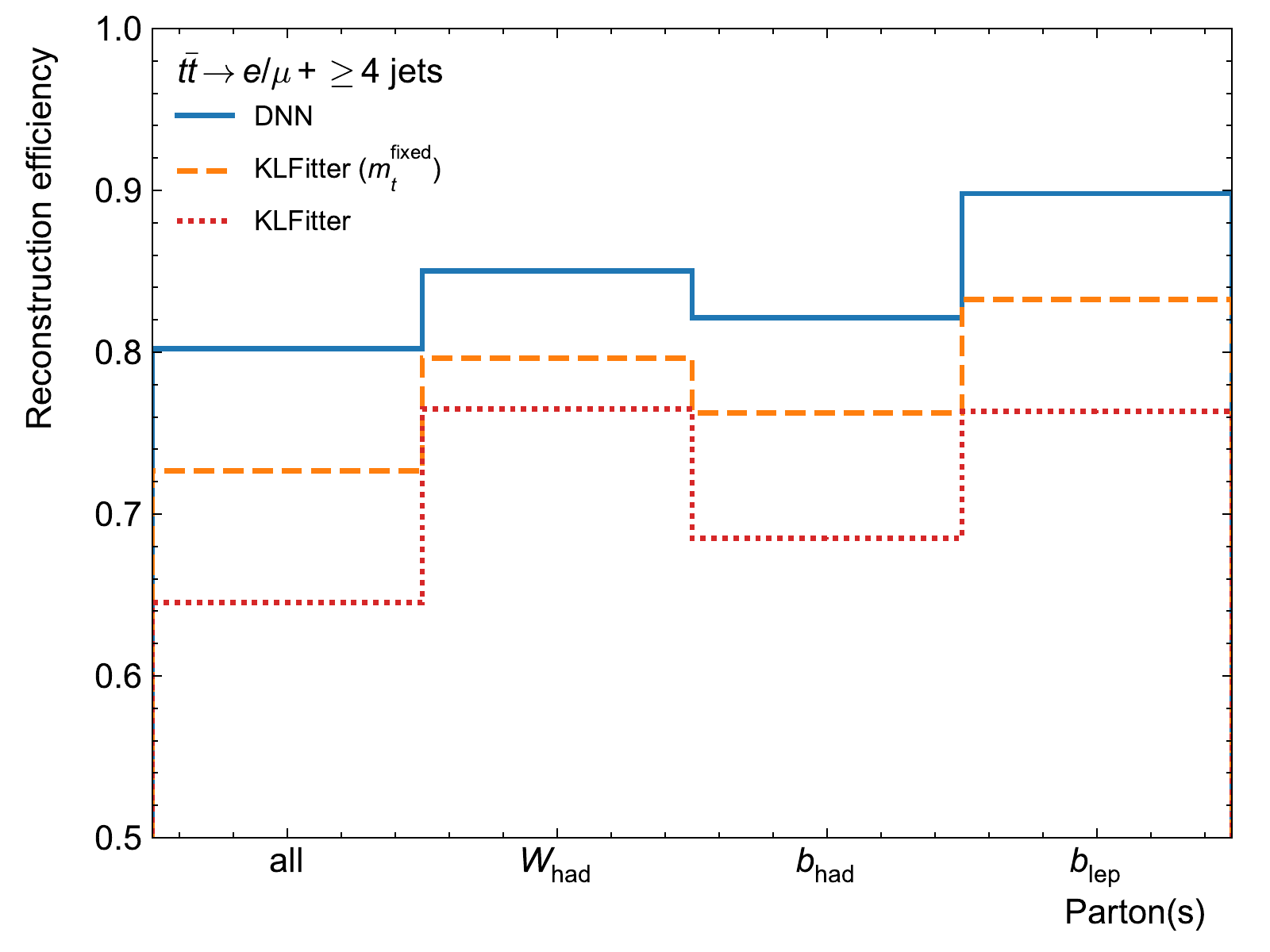}}
  \subfloat[]{\includegraphics[width=0.49\textwidth]{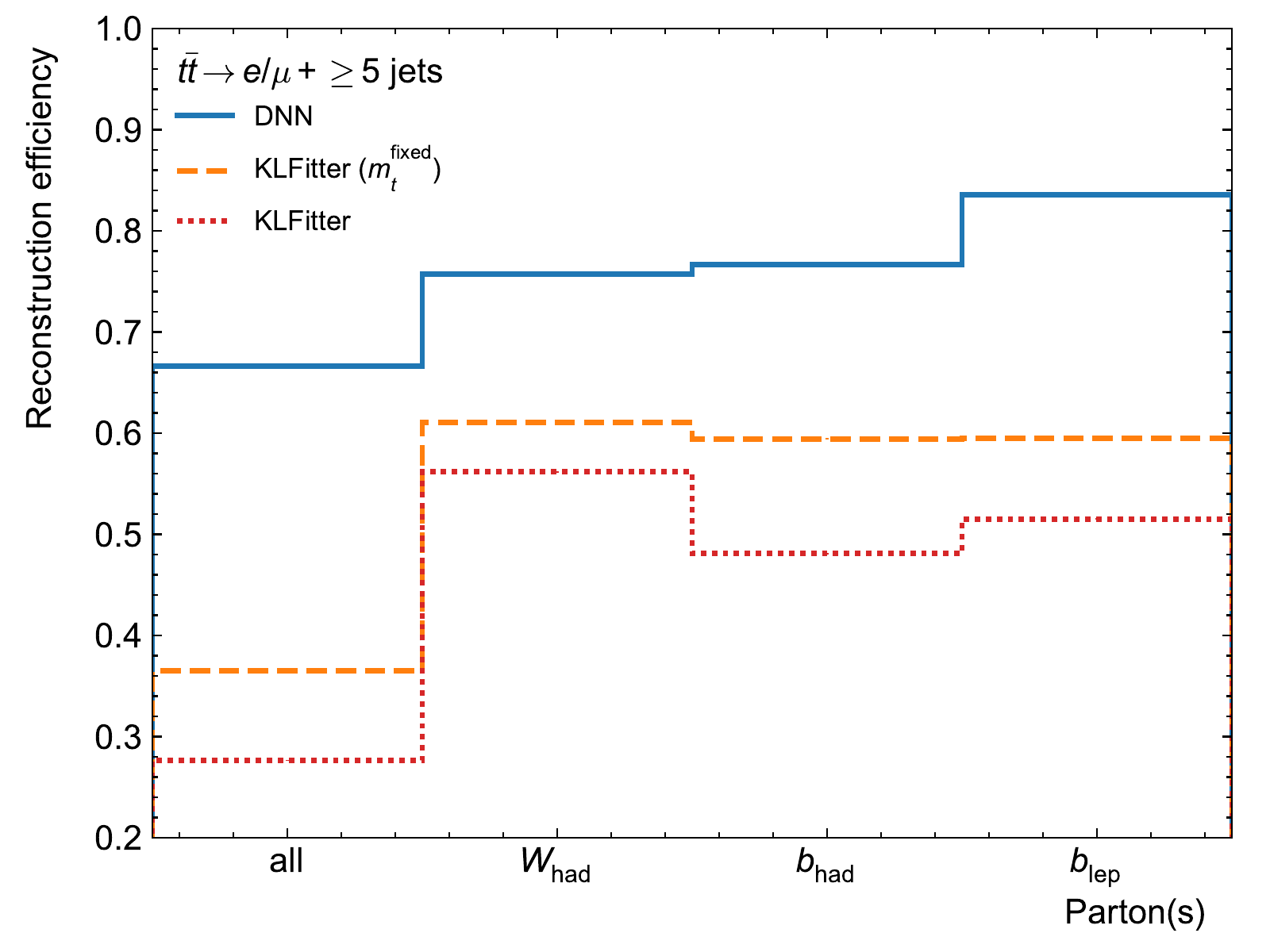}}\\
  \subfloat[]{\includegraphics[width=0.49\textwidth]{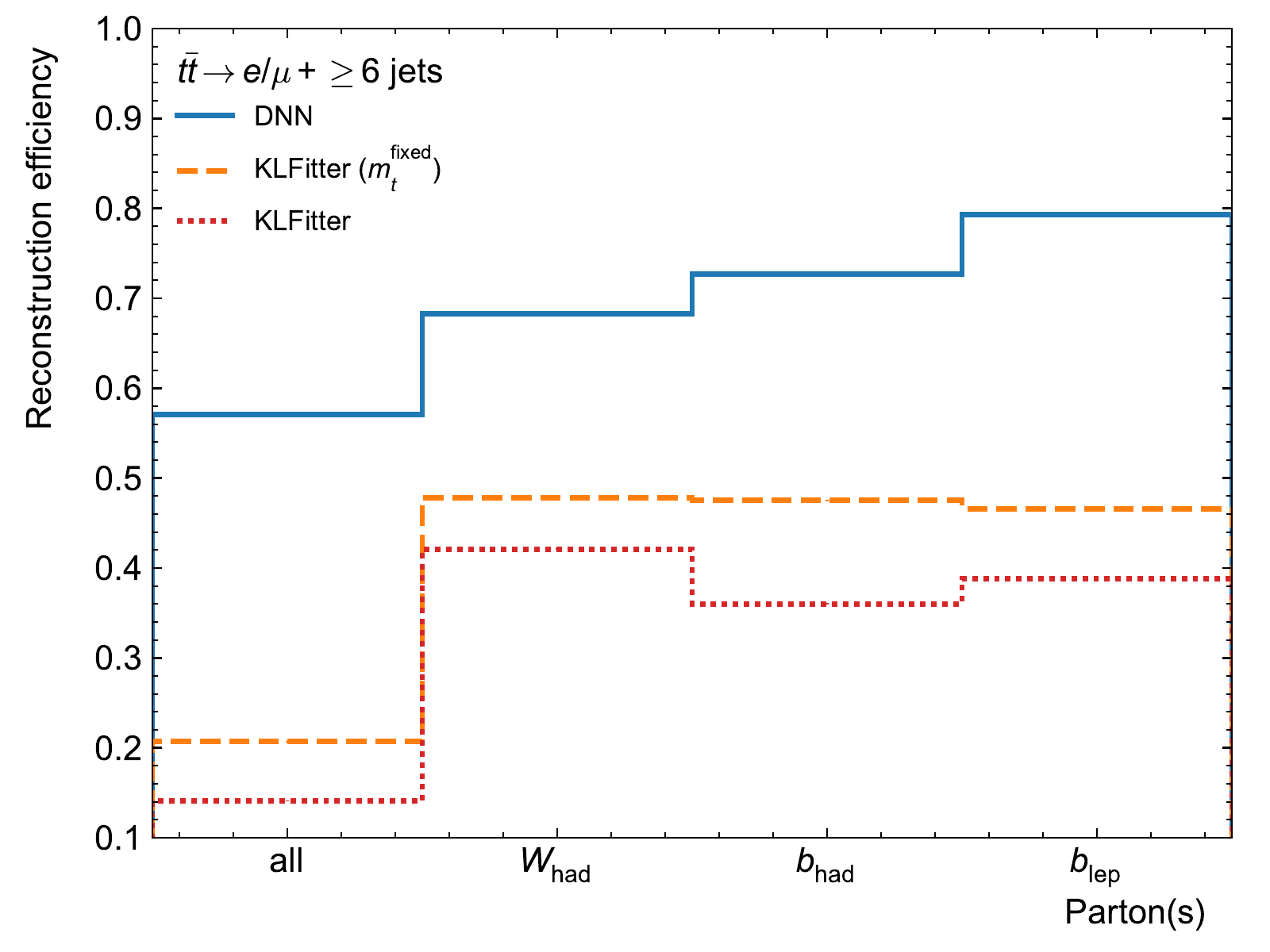}}
  \caption{Reconstruction efficiency for events with at least (a) four, (b) five and (c) six jets for all four quarks from the $\ttbar$ decay (``all''), for the two quarks from the hadronic $W$-boson decay, for the $b$-quark from the hadronic top-quark decay and for the $b$-quark from the leptonic top-quark decay. The performance of the reconstruction with the DNN is compared with two configurations of the \textsc{KLFitter} algorithm.}
  \label{fig:efficiency}
\end{figure}

\begin{table}
\centering
\caption{Reconstruction efficiency for all four quarks from the $\ttbar$ decay (``all''), for the two quarks from the hadronic $W$-boson decay, for the $b$-quark from the hadronic top-quark decay and for the $b$-quark from the leptonic top-quark decay for events with at least four, five or six jets. The performance of the reconstruction with the DNN is compared with two configurations of the \textsc{KLFitter} algorithm. The statistical uncertainties in the reconstruction efficiencies are approximately 0.1\%.}
\begin{tabular}{llcccc}
\toprule
Jet selection & Algorithm & \multicolumn{4}{c}{Reconstruction efficiency} \\
& & all & $W_{\mathrm{had}}$ & $b_{\mathrm{had}}$ & $b_{\mathrm{lep}}$ \\
\midrule
\multirow{3}{*}{$\geq 4$ jets} & DNN & 80.2\% & 85.0\% & 82.2\% & 89.8\% \\
& \textsc{KLFitter} ($m_t^{\mathrm{fixed}}$) & 72.7\% & 79.7\% & 76.2\% & 83.3\% \\
& \textsc{KLFitter} & 64.5\% & 76.5\% & 68.5\% & 76.3\% \\
\midrule
\multirow{3}{*}{$\geq 5$ jets} & DNN & 66.6\% & 75.8\% & 76.7\% & 83.6\% \\
& \textsc{KLFitter} ($m_t^{\mathrm{fixed}}$) & 36.5\% & 61.1\% & 59.4\% & 59.5\% \\
& \textsc{KLFitter} & 27.6\% & 56.2\% & 48.1\% & 51.5\% \\
\midrule
\multirow{3}{*}{$\geq 6$ jets} & DNN & 57.1\% & 68.3\% & 72.7\% & 79.3\% \\
& \textsc{KLFitter} ($m_t^{\mathrm{fixed}}$) & 20.7\% & 47.8\% & 47.5\% & 46.6\% \\
& \textsc{KLFitter} & 14.1\% & 42.1\% & 36.0\% & 38.8\% \\
\bottomrule
\end{tabular}
\label{tab:recoeff}
\end{table}

In Figure~\ref{fig:efficiency}, the performance of the DNN is compared to the performance achieved with the \textsc{KLFitter} configurations using unambiguously matched events, as defined in Section~\ref{sec:samples_selection} and used for the DNN training. The values from the figure are also displayed in Table~\ref{tab:recoeff}. In these events, the reconstruction efficiency for the correct assignment of jets to all four quarks from the $\ttbar$ decay shows a better performance of the DNN compared to \textsc{KLFitter}, where---as expected---the \textsc{KLFitter} configuration with the fixed top-quark mass shows a better performance than the floating-mass configuration~\cite{Erdmann:2013rxa}. The reconstruction efficiency is in addition compared for the correct assignment of only a part of the event. Also in the cases of the correct assignment of jets to the two quarks from the hadronic $W$-boson decay, to the $b$-quark from the hadronic top-quark decay and to the $b$-quark from the leptonic top-quark decay, the DNN outperforms the \textsc{KLFitter} algorithm. The improvement is particularly strong in the case of events with more than four jets, which are especially difficult to reconstruct, because in these events at least one additional jet is present---with respect to the leading-order picture of lepton+jets $\ttbar$ events with four quarks in the final state. Such events are a challenge for the \textsc{KLFitter} algorithm, which is constructed based on this leading-order picture. The DNN, however, receives also five- and six-jet events during the training, can learn the properties of jets originating from additional radiation and is hence able to identify the correct jet permutation with a higher probability in these cases.

\begin{figure}[p]
  \centering
  \subfloat[]{\includegraphics[width=0.49\textwidth]{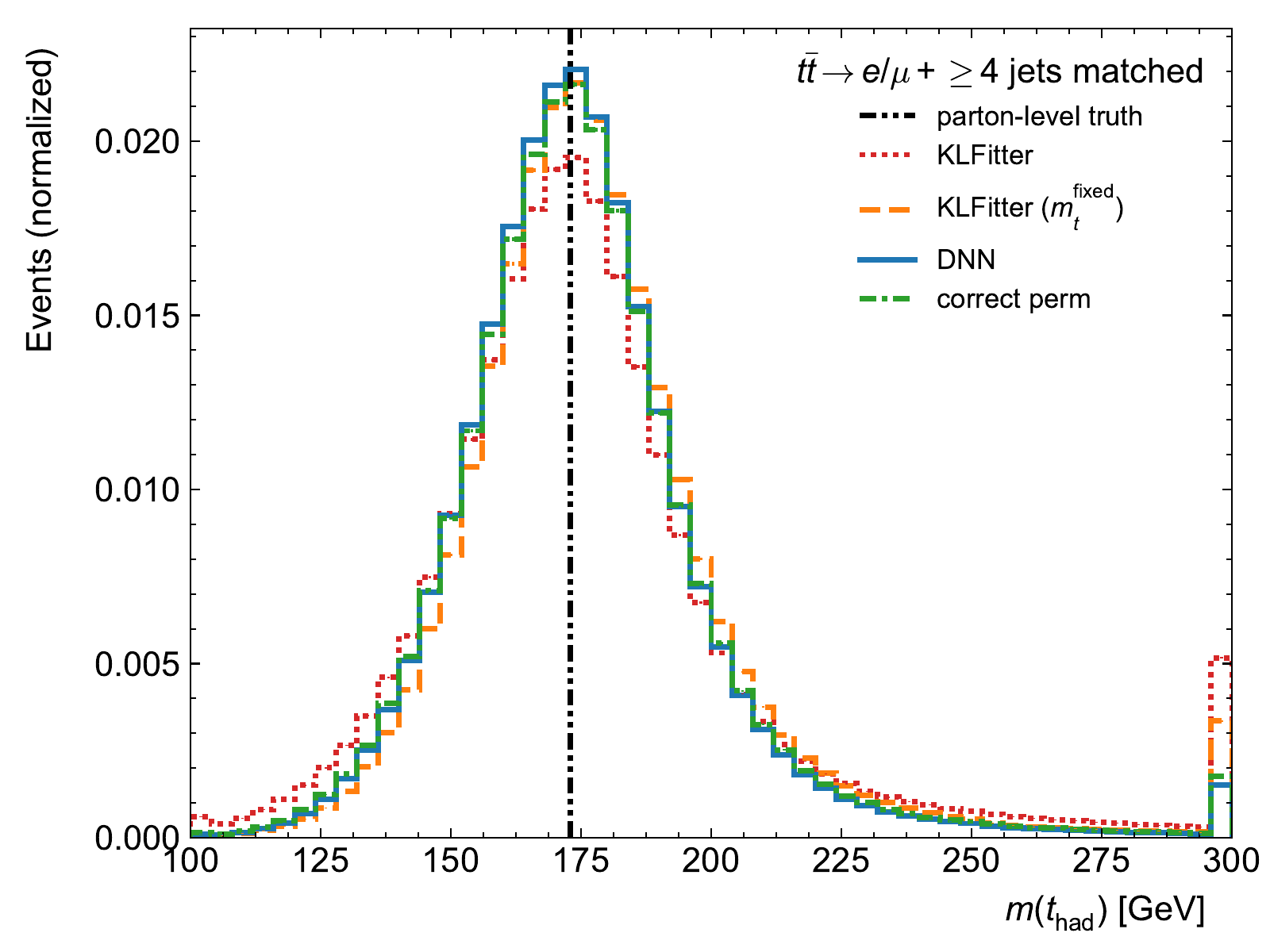}}
  \subfloat[]{\includegraphics[width=0.49\textwidth]{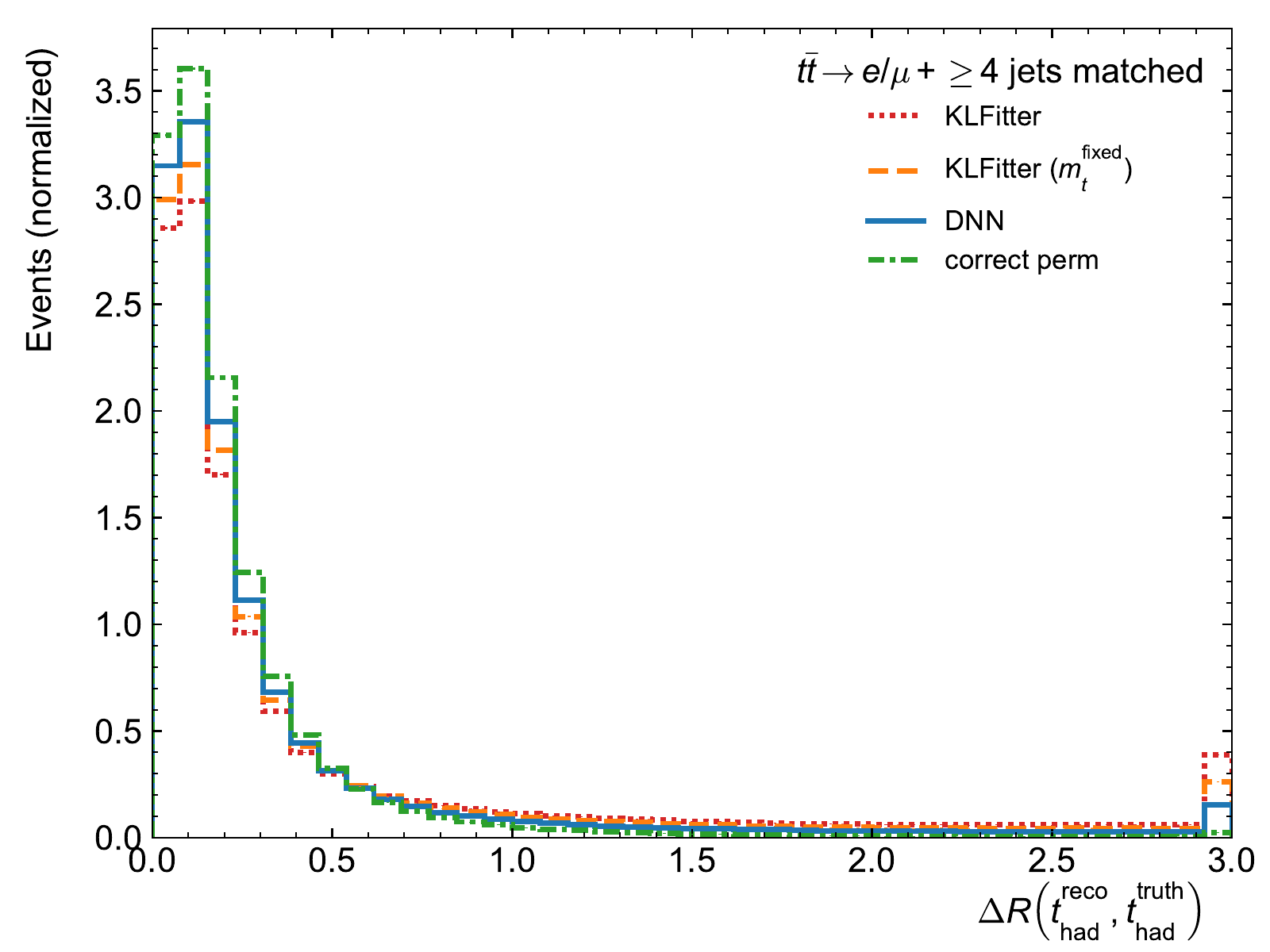}}\\
  \subfloat[]{\includegraphics[width=0.49\textwidth]{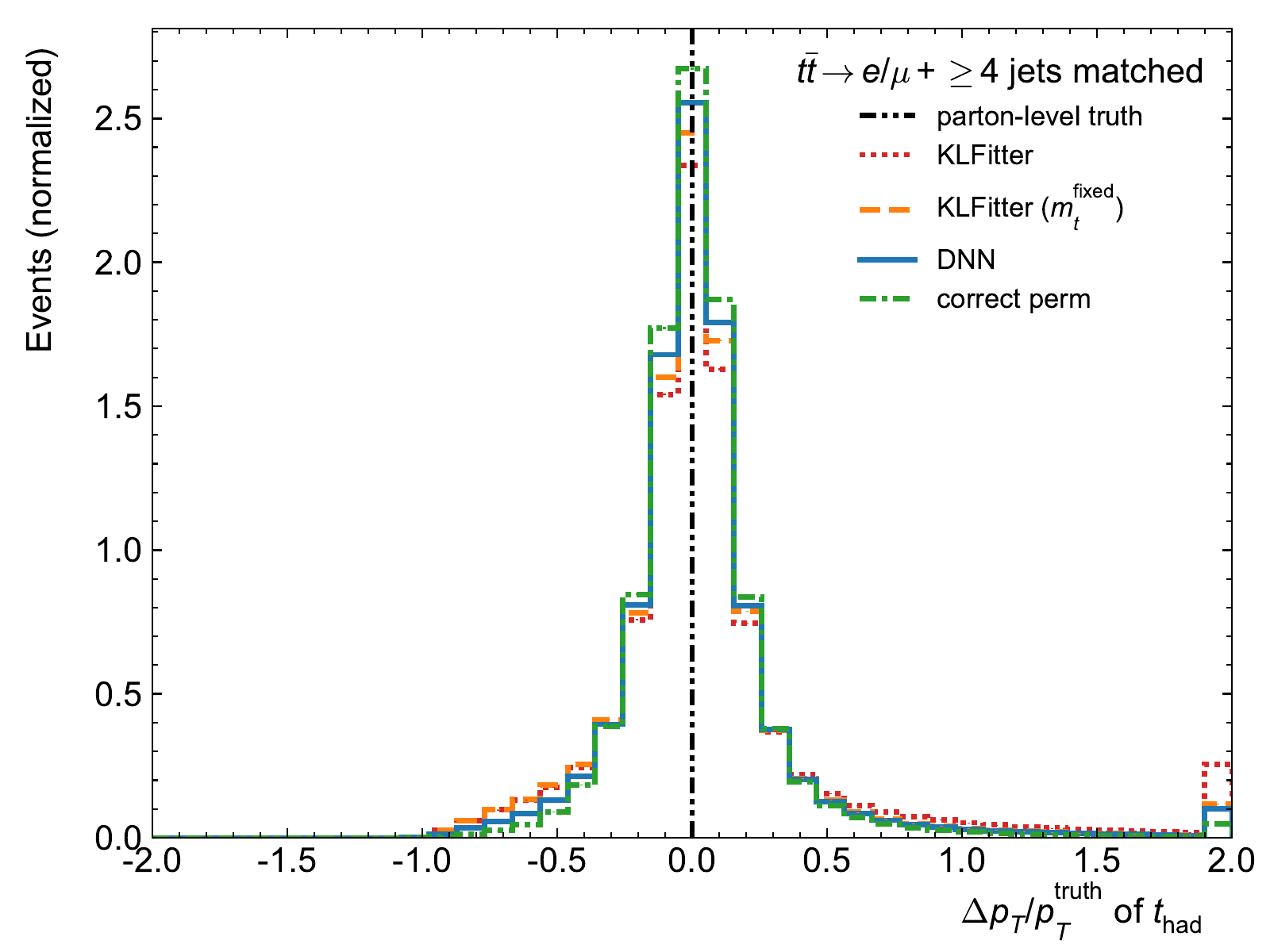}}
  \caption{Distributions for events with at least four jets in unambiguously matched events (``matched'') of (a) the reconstructed mass of the hadronically-decaying top quark, (b) its difference in direction with respect to the true top quark in $\eta$-$\Phi$ space and (c) the relative difference of the \pt\ ($\Delta \pt/p_{\mathrm{T}}^{\mathrm{truth}}$ with $\Delta \pt = p_{\mathrm{T}}^{\mathrm{predicted}}-p_{\mathrm{T}}^{\mathrm{truth}}$) with respect to the true top-quark's \pt. Also shown is the distribution of the correct permutation (``correct perm''). In the case of the mass, the top-quark mass used in the Monte Carlo simulation is indicated by a vertical line. The performance of the reconstruction with the DNN is compared with two configurations of the \textsc{KLFitter} algorithm.}
  \label{fig:kinematics_4j_matched}
\end{figure}

\begin{figure}[p]
  \centering
  \subfloat[]{\includegraphics[width=0.49\textwidth]{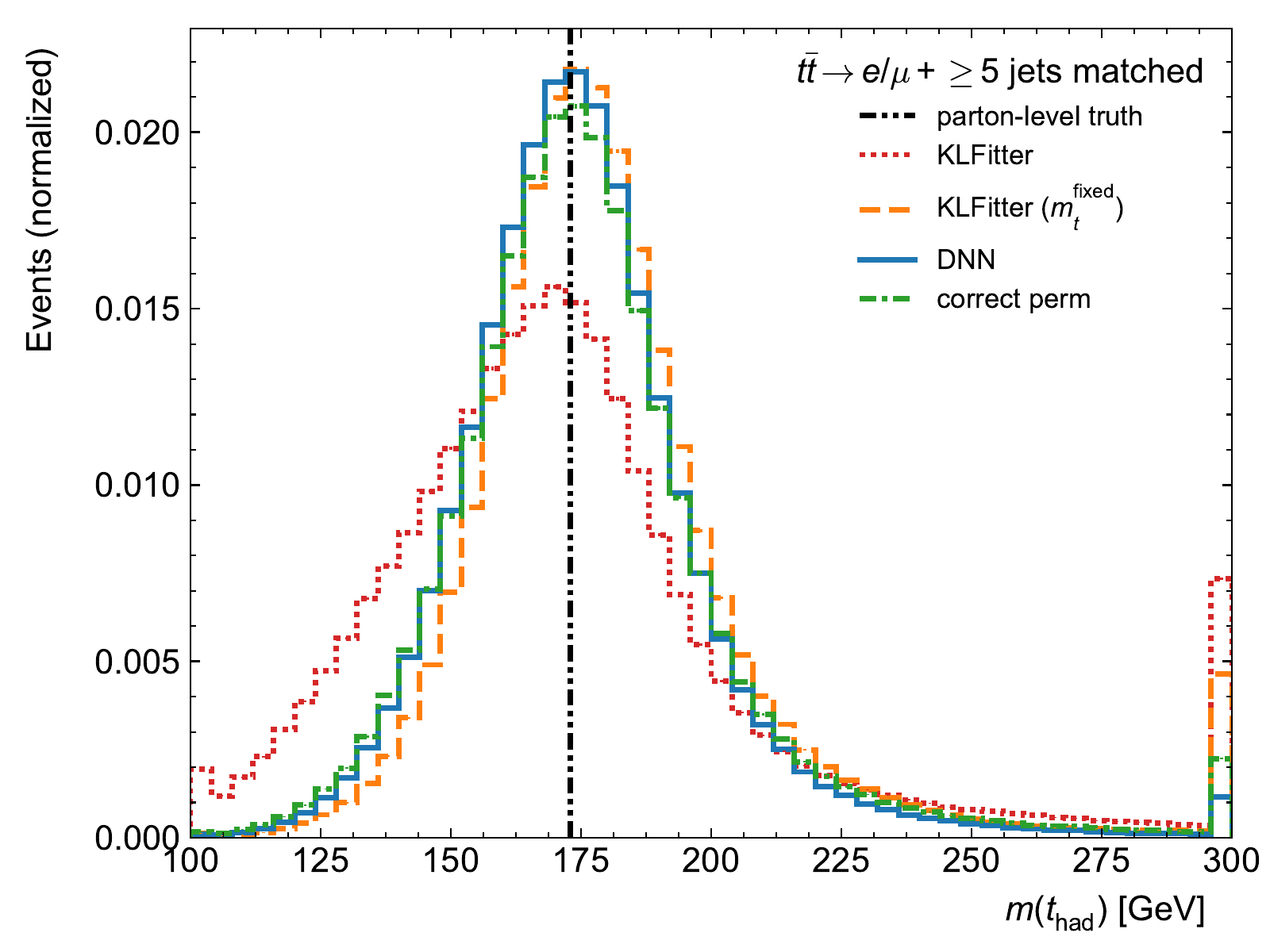}}
  \subfloat[]{\includegraphics[width=0.49\textwidth]{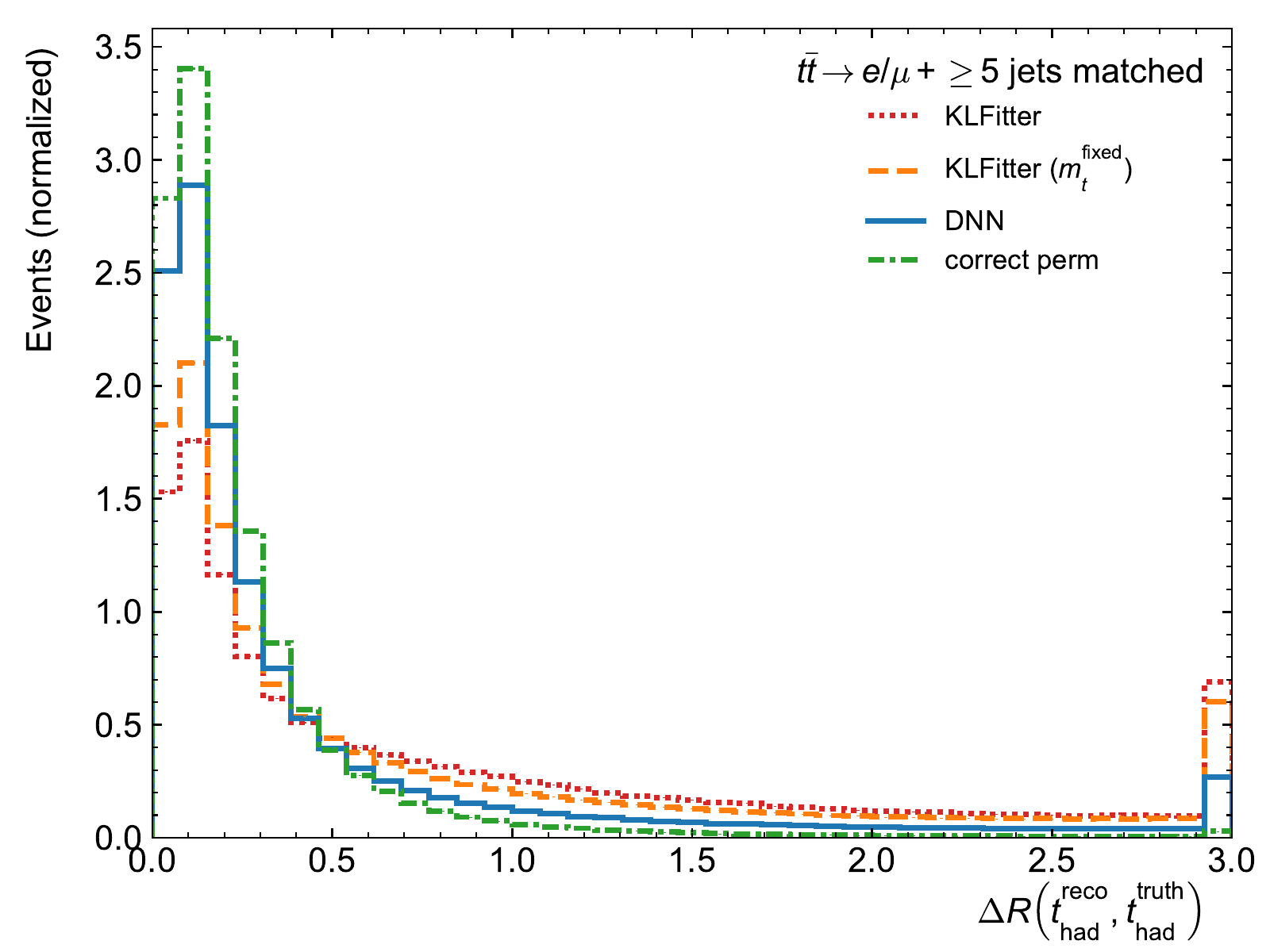}}\\
  \subfloat[]{\includegraphics[width=0.49\textwidth]{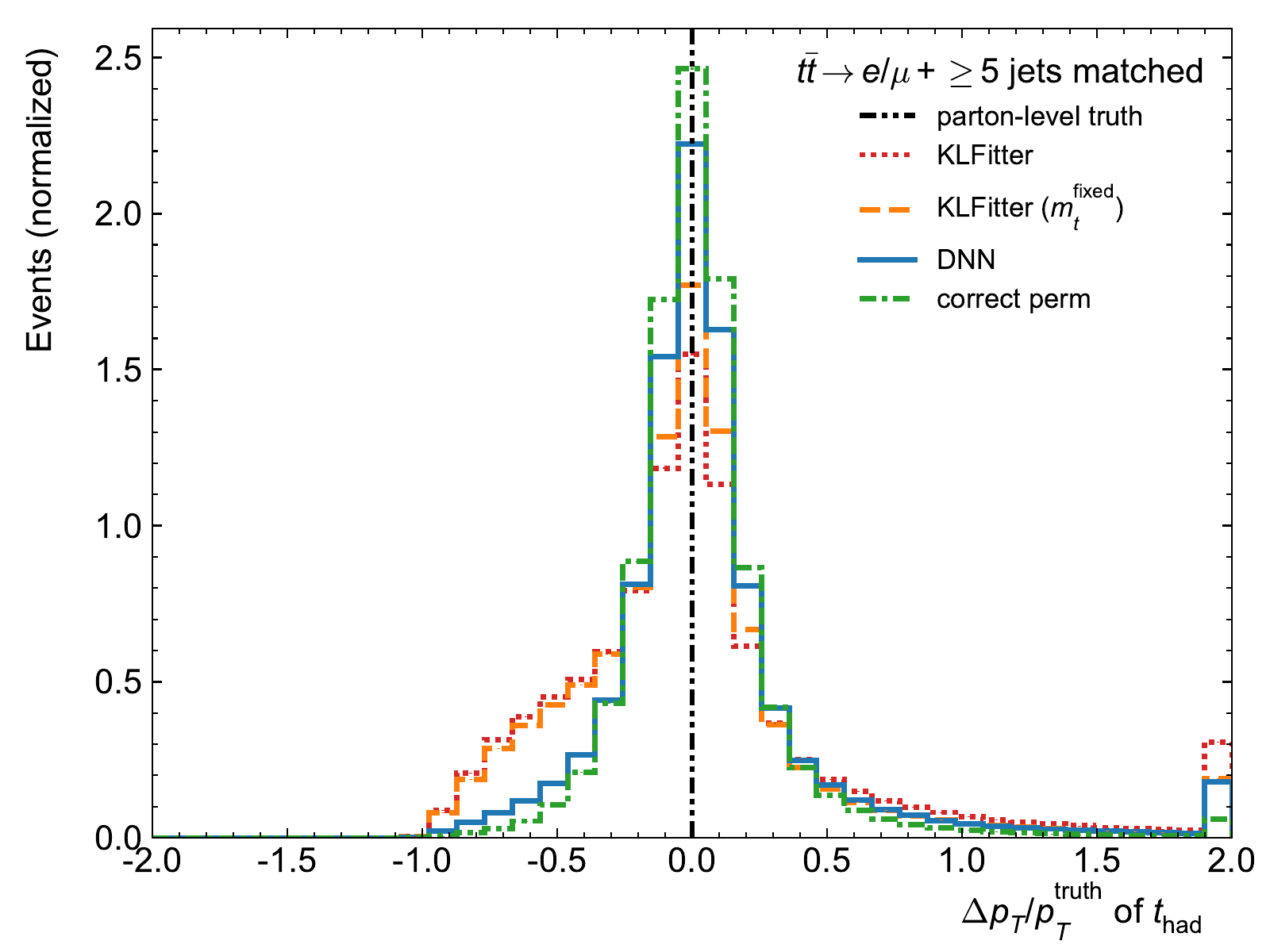}}
  \caption{Distributions for events with at least five jets in unambiguously matched events (``matched'') of (a) the reconstructed mass of the hadronically-decaying top quark, (b) its difference in direction with respect to the true top quark in $\eta$-$\Phi$ space and (c) the relative difference of the \pt\ ($\Delta \pt/p_{\mathrm{T}}^{\mathrm{truth}}$ with $\Delta \pt = p_{\mathrm{T}}^{\mathrm{predicted}}-p_{\mathrm{T}}^{\mathrm{truth}}$) with respect to the true top-quark's \pt. Also shown is the distribution of the correct permutation (``correct perm''). In the case of the mass, the top-quark mass used in the Monte Carlo simulation is indicated by a vertical line. The performance of the reconstruction with the DNN is compared with two configurations of the \textsc{KLFitter} algorithm.}
  \label{fig:kinematics_5j_matched}
\end{figure}

\begin{figure}[p]
  \centering
  \subfloat[]{\includegraphics[width=0.49\textwidth]{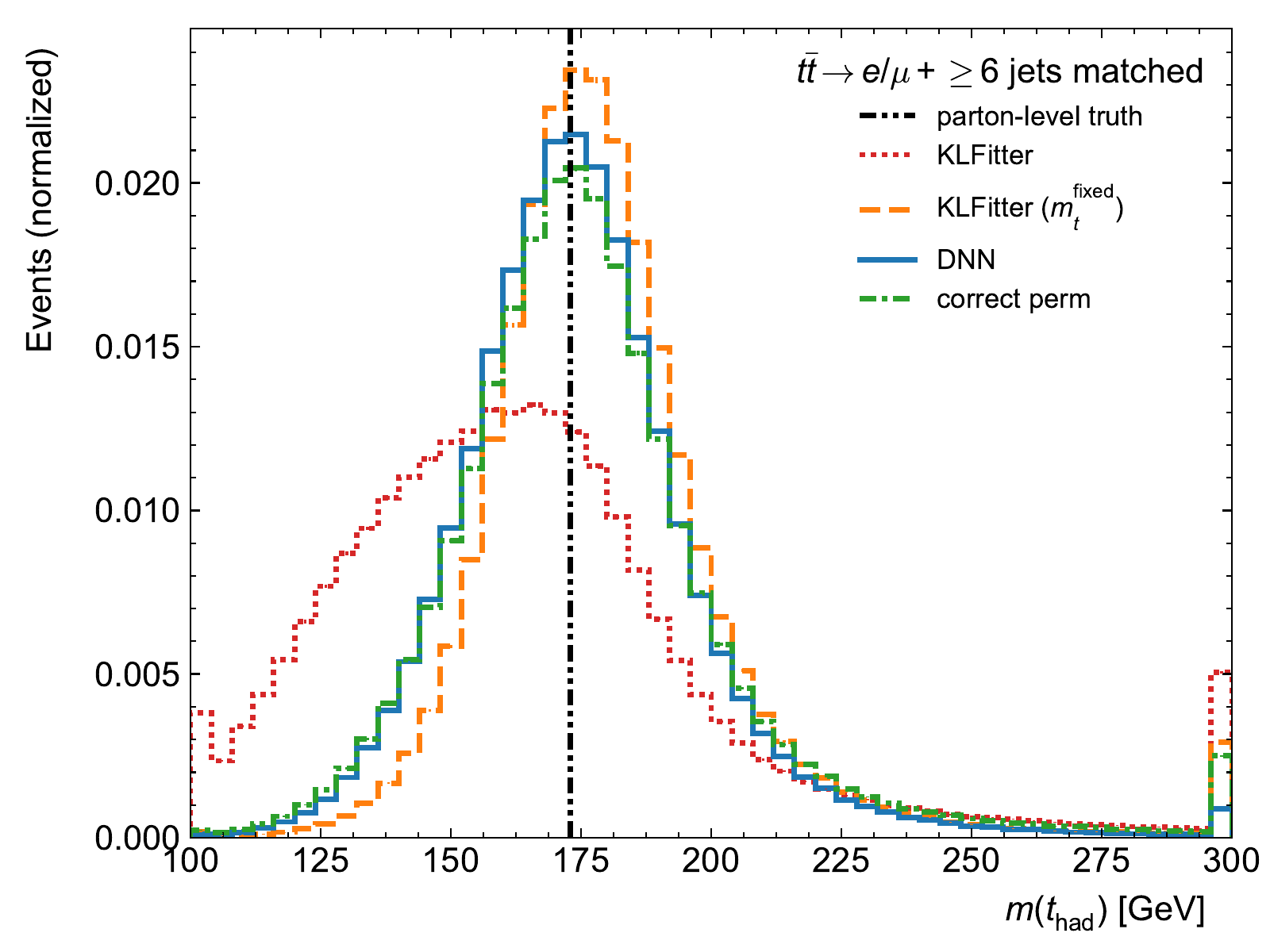}}
  \subfloat[]{\includegraphics[width=0.49\textwidth]{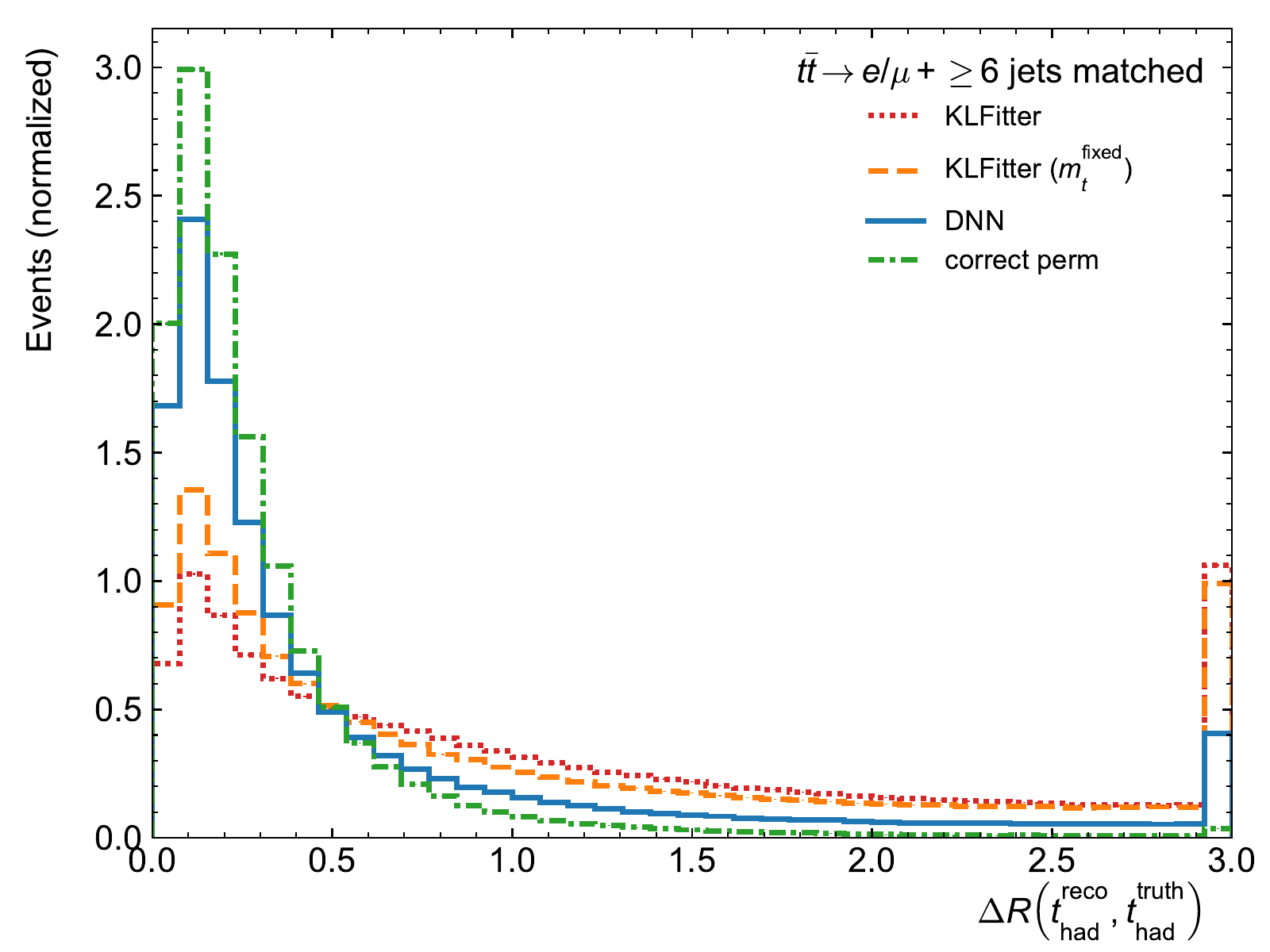}}\\
  \subfloat[]{\includegraphics[width=0.49\textwidth]{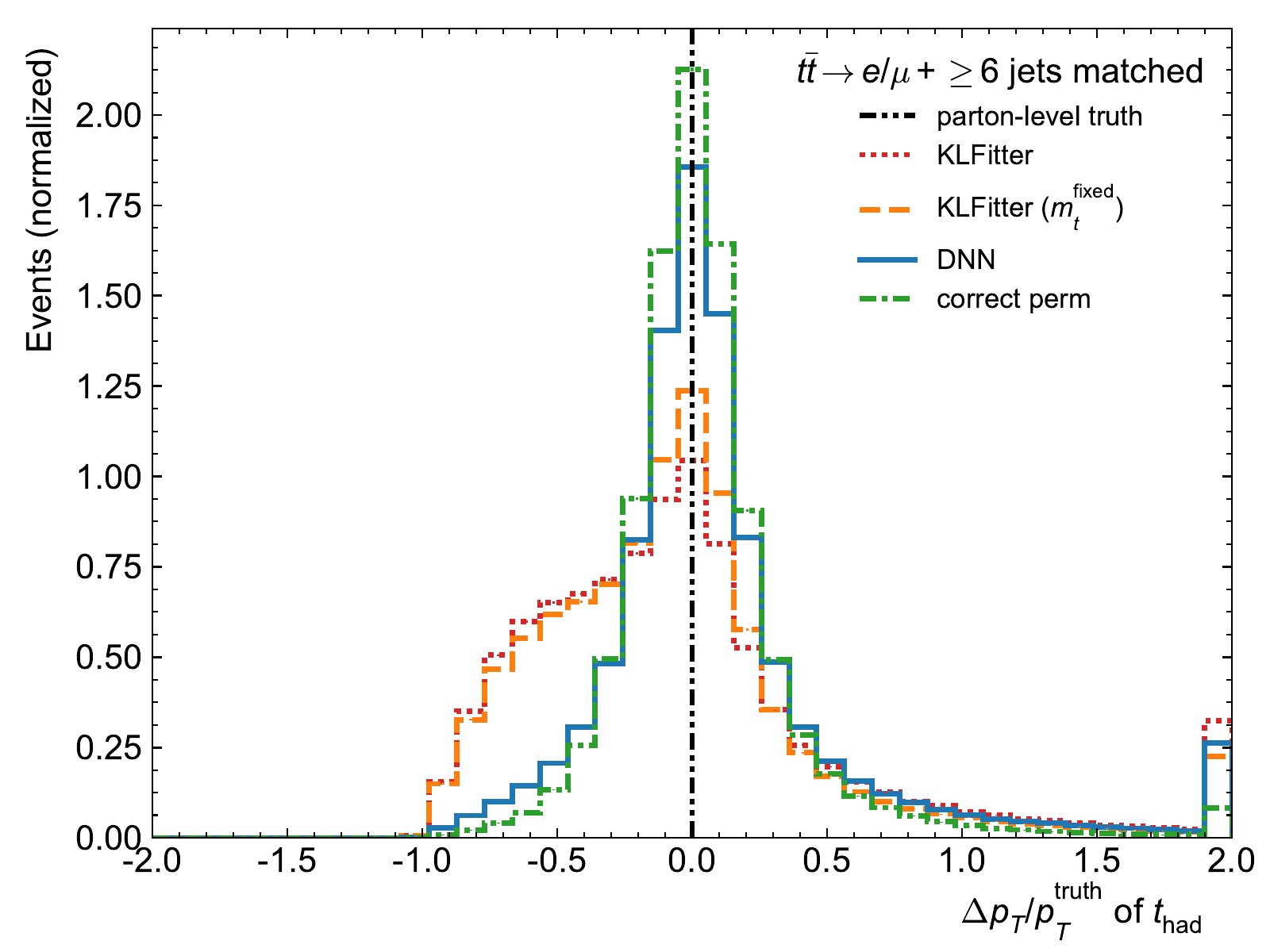}}
  \caption{Distributions for events with at least six jets in unambiguously matched events (``matched'') of (a) the reconstructed mass of the hadronically-decaying top quark, (b) its difference in direction with respect to the true top quark in $\eta$-$\Phi$ space and (c) the relative difference of the \pt\ ($\Delta \pt/p_{\mathrm{T}}^{\mathrm{truth}}$ with $\Delta \pt = p_{\mathrm{T}}^{\mathrm{predicted}}-p_{\mathrm{T}}^{\mathrm{truth}}$) with respect to the true top-quark's \pt. Also shown is the distribution of the correct permutation (``correct perm''). In the case of the mass, the top-quark mass used in the Monte Carlo simulation is indicated by a vertical line. The performance of the reconstruction with the DNN is compared with two configurations of the \textsc{KLFitter} algorithm.}
  \label{fig:kinematics_6j_matched}
\end{figure}

In Figures~\ref{fig:kinematics_4j_matched}--\ref{fig:kinematics_6j_matched}, distributions of the reconstructed mass of the top quark that decays hadronically, its difference in direction with respect to the top quark at parton level (``true top quark'') in $\eta$-$\Phi$ space and the relative difference of the \pt\ with respect to the true top-quark's \pt\ are shown for the four-, five- and six-jet selections. In all figures, only events are shown that are unambiguously matched, i.e.\ the type of events that is used for training the DNN and where a correct permutation is well-defined. The distribution of the correct permutation is also shown, corresponding to a perfect reconstruction algorithm.

The high reconstruction efficiency of the DNN is reflected in its mass distribution being close to the distribution of the correct permutation for all jet selections (Figures~\ref{fig:kinematics_4j_matched}(a)--\ref{fig:kinematics_6j_matched}(a)). \textsc{KLFitter} in the fixed-mass mode shows a mass distribution that is close to the distribution of the correct permutation for the four-jet selection and also shows narrow mass peaks for the five- and six-jet selections, which is expected as the algorithm uses the true top-quark mass as the central value of the top-quark-mass Breit-Wigner functions. However, the mass distributions are slightly shifted to higher values, especially for the five- and six-jet selections, because with increasing freedom in the choice of permutations, \textsc{KLFitter} may choose a wrong permutations, in which at least one chosen jet does not originate from the hadronically-decaying top-quark. This is also indicated by the fact that the mass peak of the distribution for correct permutations is less pronounced than the one reconstructed by \textsc{KLFitter} for the five- and six-jet selections.

The reconstruction of the direction of the hadronically-decaying top quark with respect to the true top quark's distribution is better in the case of the reconstruction with the DNN compared to both \textsc{KLFitter} modes in the four-jet case, with the DNN distribution being closest to the distribution of the correct permutation (Figure~\ref{fig:kinematics_4j_matched}(b)). The same holds for the relative difference of the \pt\ distribution with respect to the true top quark's \pt\ (Figure~\ref{fig:kinematics_4j_matched}(c)). The improvement due to the reconstruction with the DNN is even much stronger for the selections with at least five or at least six jets (Figures~\ref{fig:kinematics_5j_matched}(b,c) and~\ref{fig:kinematics_6j_matched}(b,c)) than seen in the four-jet case. As expected, \textsc{KLFitter} in the floating-mass mode results in a worse performance in all of these distributions than \textsc{KLFitter} in the fixed-mass mode.

These results indicate that the DNN is indeed able to learn the kinematics and $b$-tagging characteristics of the \ttbar\ decay chain (including additional radiation) and identify the correct permutation by exploring this information. As expected from the higher reconstruction efficiencies, the DNN outperforms the two \textsc{KLFitter} modes also in terms of the studied kinematic variables.

In the training of the DNN and in the studies presented so far only unambiguously matched events were used. However, it is important to evaluate the reconstruction with the DNN not only in this particular phase space, but using all events that pass the selection. In unmatched events a true permutation is not well-defined and we evaluate the performance of the reconstruction based on top-quark kinematic variables that are built using the predicted permutation. A better reconstruction algorithm, i.e.\ a better jet-to-parton assignment, is expected to result in a better reconstruction of the kinematics of the top quarks, which are constructed by adding the four-vectors of the jets that are assigned by the reconstruction algorithm.

\begin{figure}[p]
  \centering
  \subfloat[]{\includegraphics[width=0.49\textwidth]{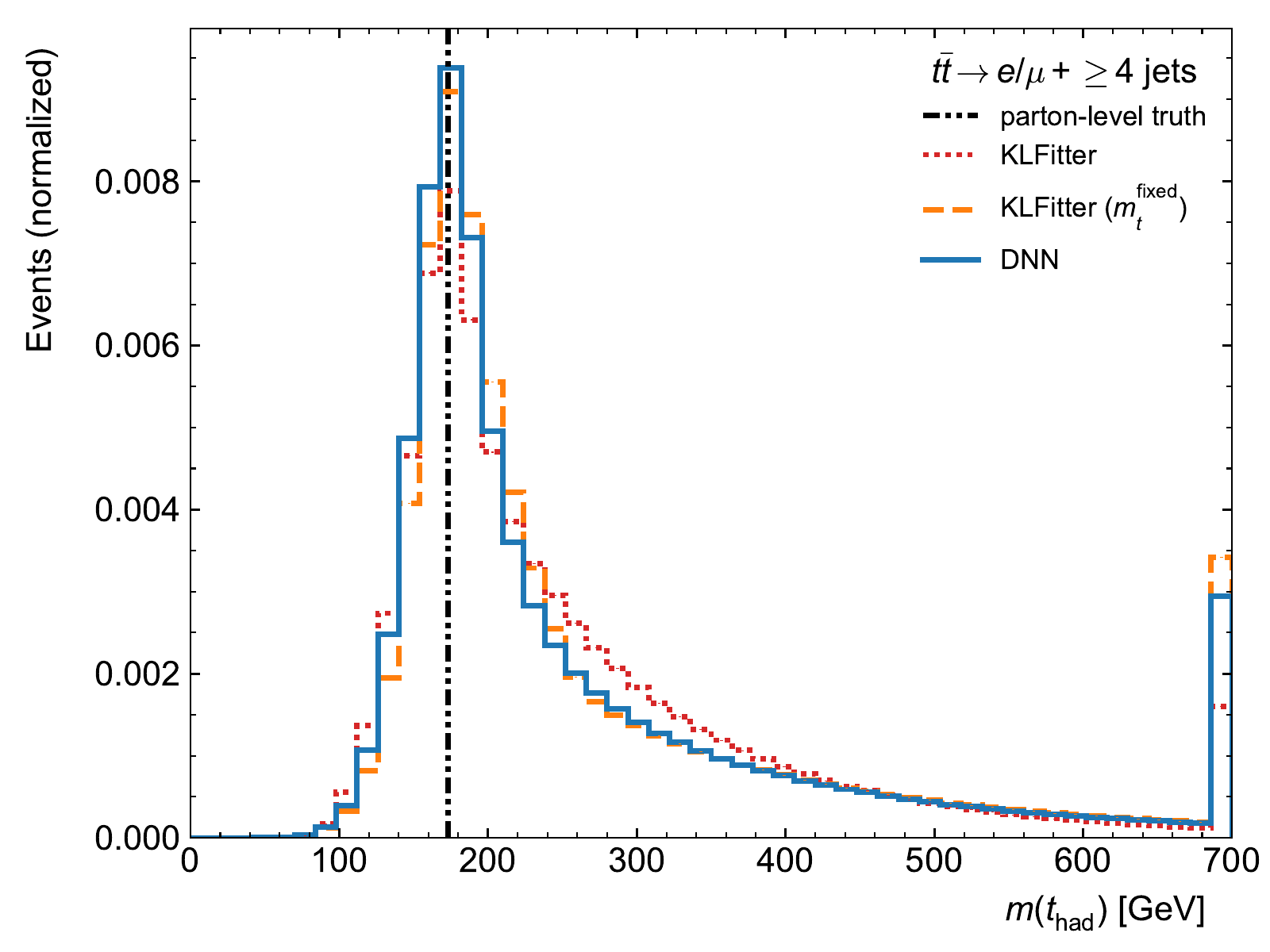}}
  \subfloat[]{\includegraphics[width=0.49\textwidth]{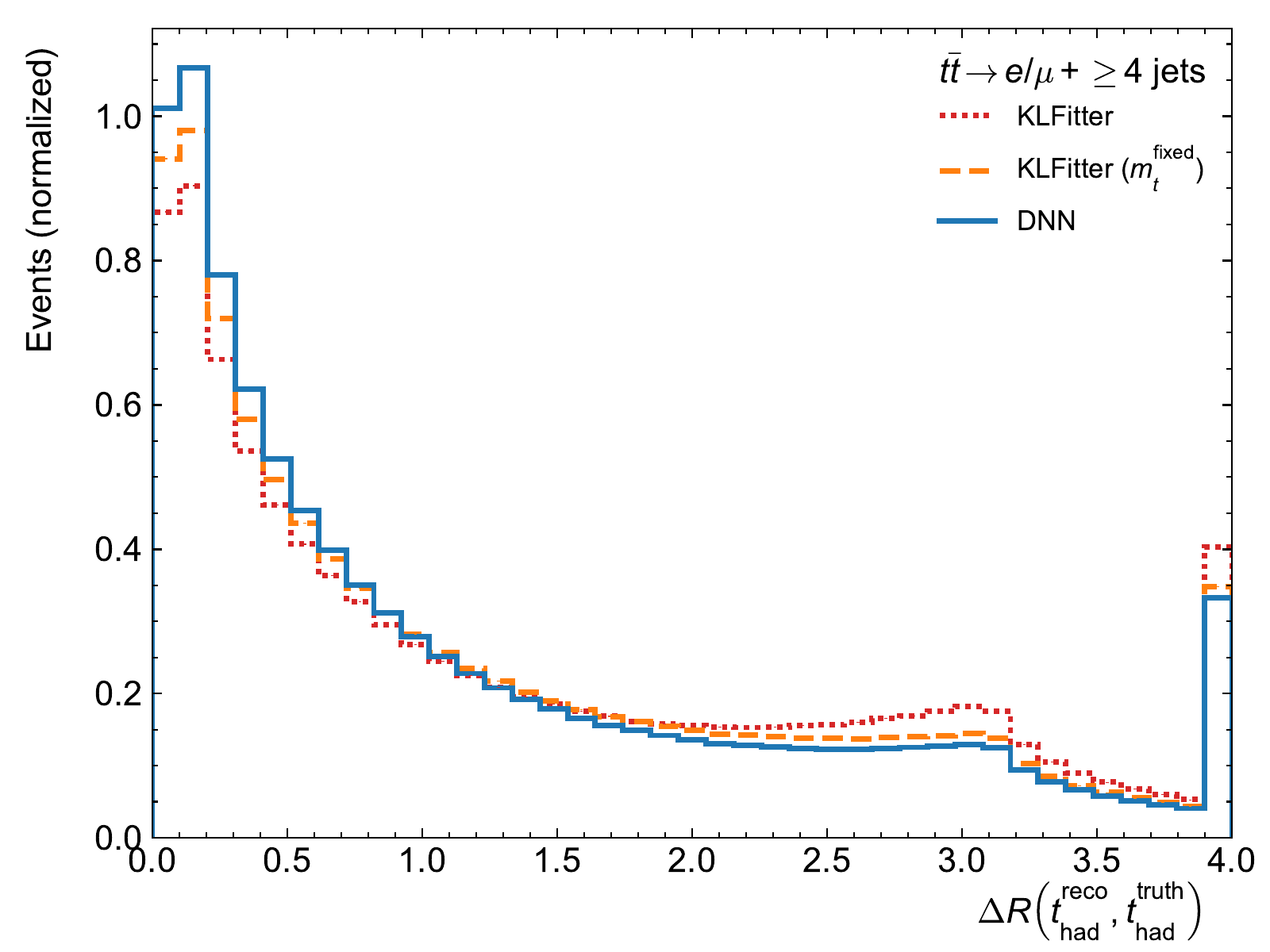}}\\
  \subfloat[]{\includegraphics[width=0.49\textwidth]{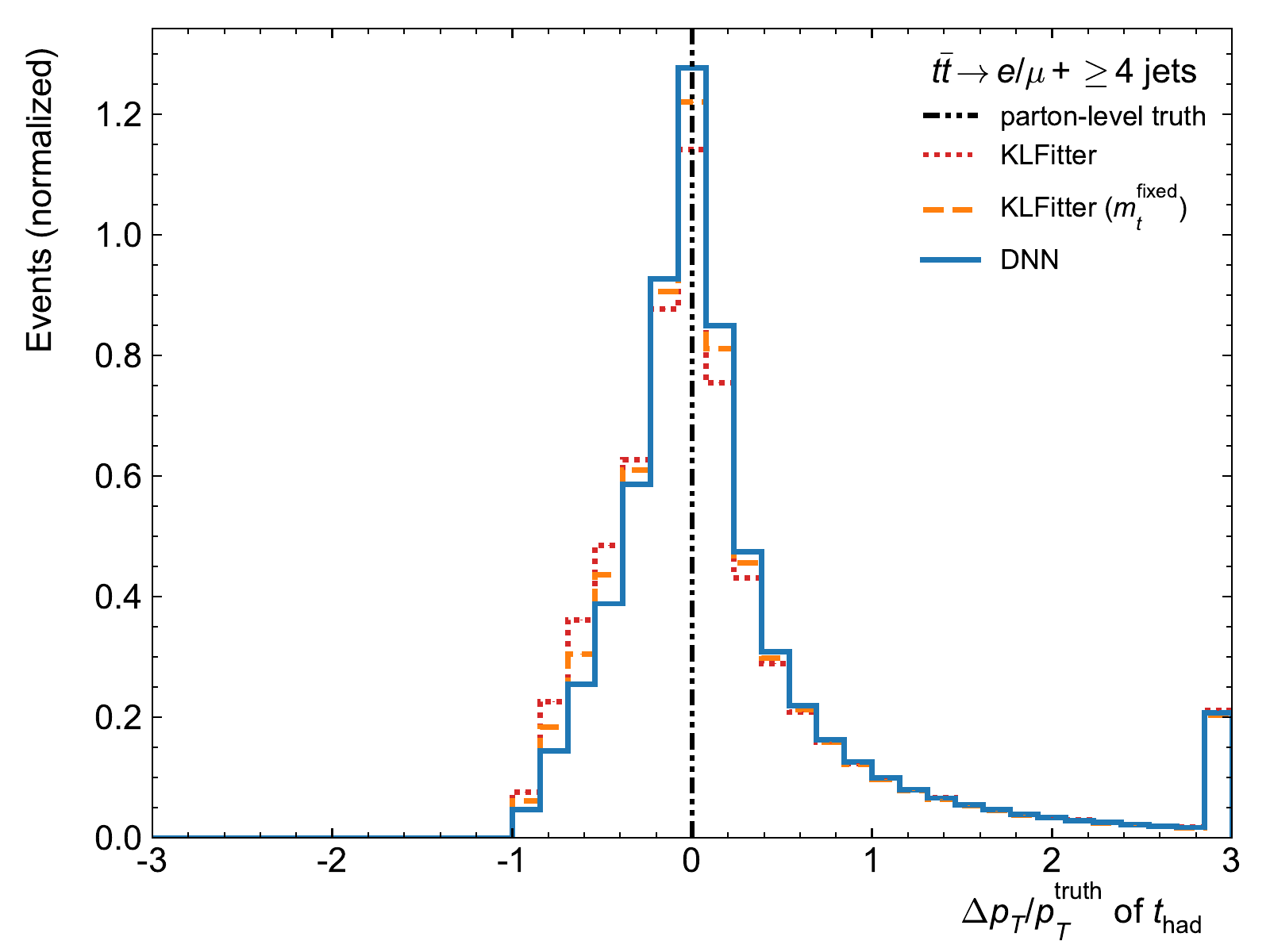}}
  \caption{Distributions for events with at least four jets of (a) the reconstructed mass of the hadronically-decaying top quark, (b) its difference in direction with respect to the true top quark in $\eta$-$\Phi$ space and (c) the relative difference of the \pt\ ($\Delta \pt/p_{\mathrm{T}}^{\mathrm{truth}}$ with $\Delta \pt = p_{\mathrm{T}}^{\mathrm{predicted}}-p_{\mathrm{T}}^{\mathrm{truth}}$) with respect to the true top-quark's \pt. In the case of the mass, the top-quark mass used in the Monte Carlo simulation is indicated by a vertical line. The performance of the reconstruction with the DNN is compared with two configurations of the \textsc{KLFitter} algorithm.}
  \label{fig:kinematics_4j}
\end{figure}

\begin{figure}[p]
  \centering
  \subfloat[]{\includegraphics[width=0.49\textwidth]{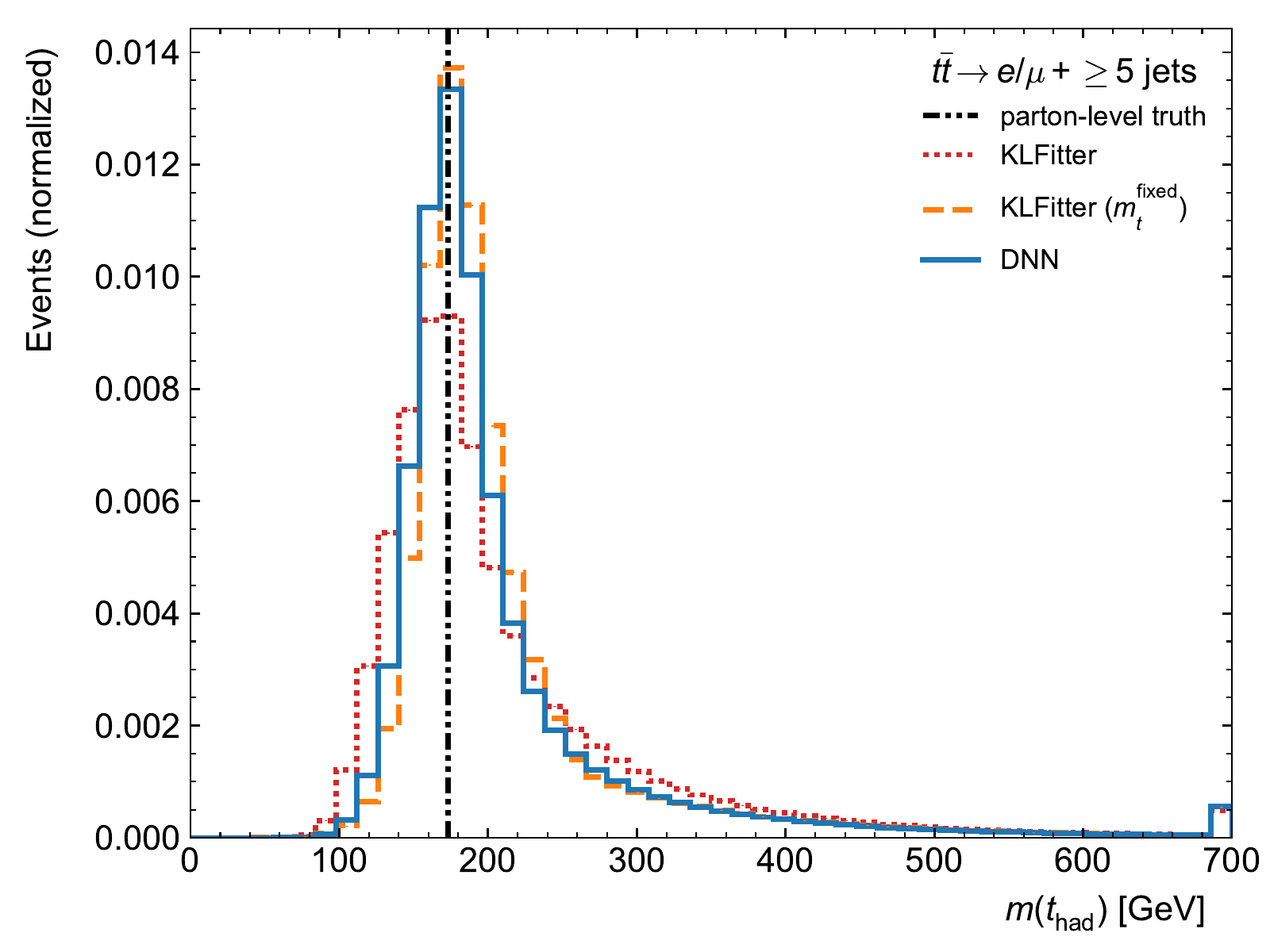}}
  \subfloat[]{\includegraphics[width=0.49\textwidth]{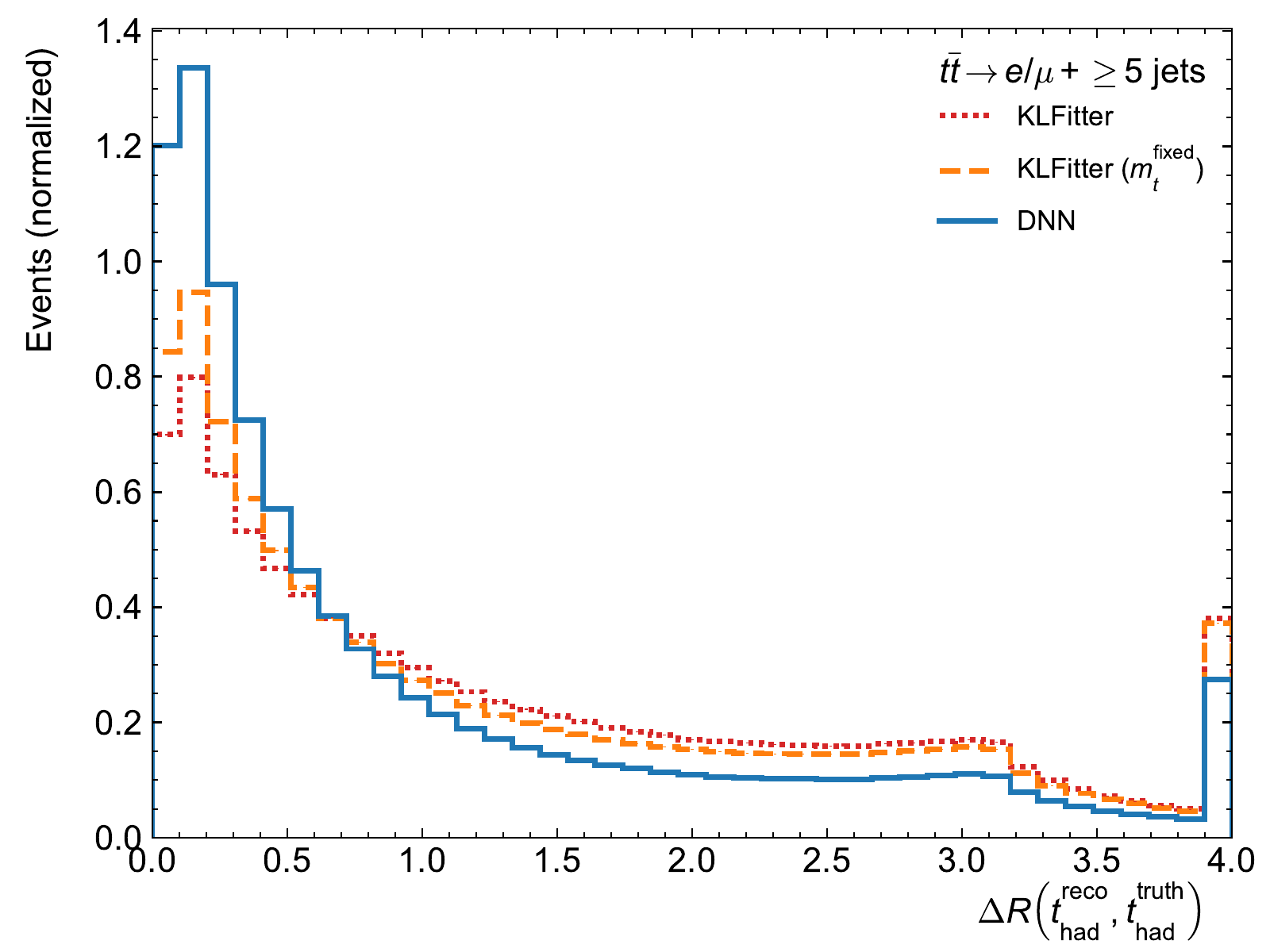}}\\
  \subfloat[]{\includegraphics[width=0.49\textwidth]{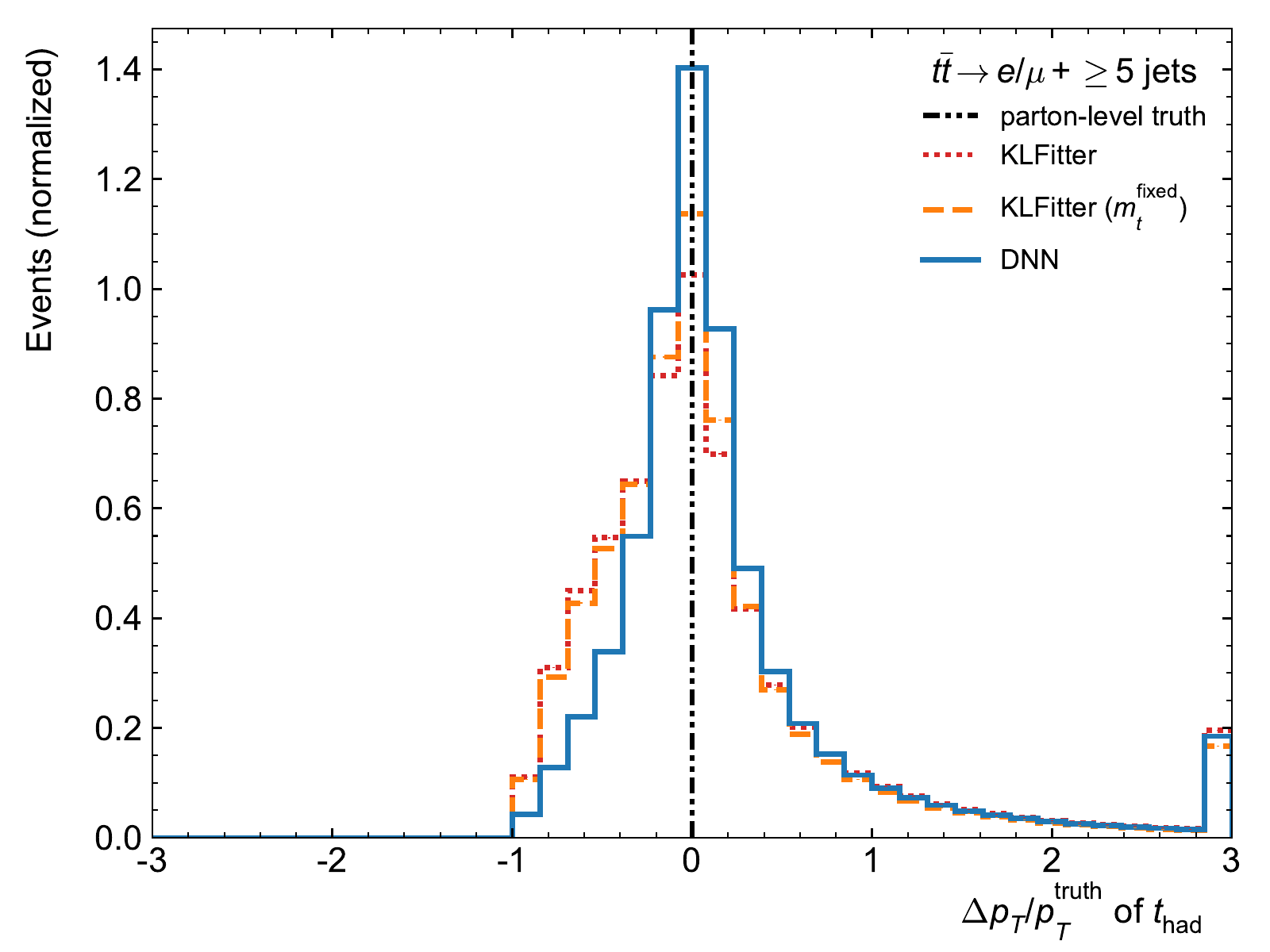}}
  \caption{Distributions for events with at least five jets of (a) the reconstructed mass of the hadronically-decaying top quark, (b) its difference in direction with respect to the true top quark in $\eta$-$\Phi$ space and (c) the relative difference of the \pt\ ($\Delta \pt/p_{\mathrm{T}}^{\mathrm{truth}}$ with $\Delta \pt = p_{\mathrm{T}}^{\mathrm{predicted}}-p_{\mathrm{T}}^{\mathrm{truth}}$) with respect to the true top-quark's \pt. In the case of the mass, the top-quark mass used in the Monte Carlo simulation is indicated by a vertical line. The performance of the reconstruction with the DNN is compared with two configurations of the \textsc{KLFitter} algorithm.}
  \label{fig:kinematics_5j}
\end{figure}

\begin{figure}[p]
  \centering
  \subfloat[]{\includegraphics[width=0.49\textwidth]{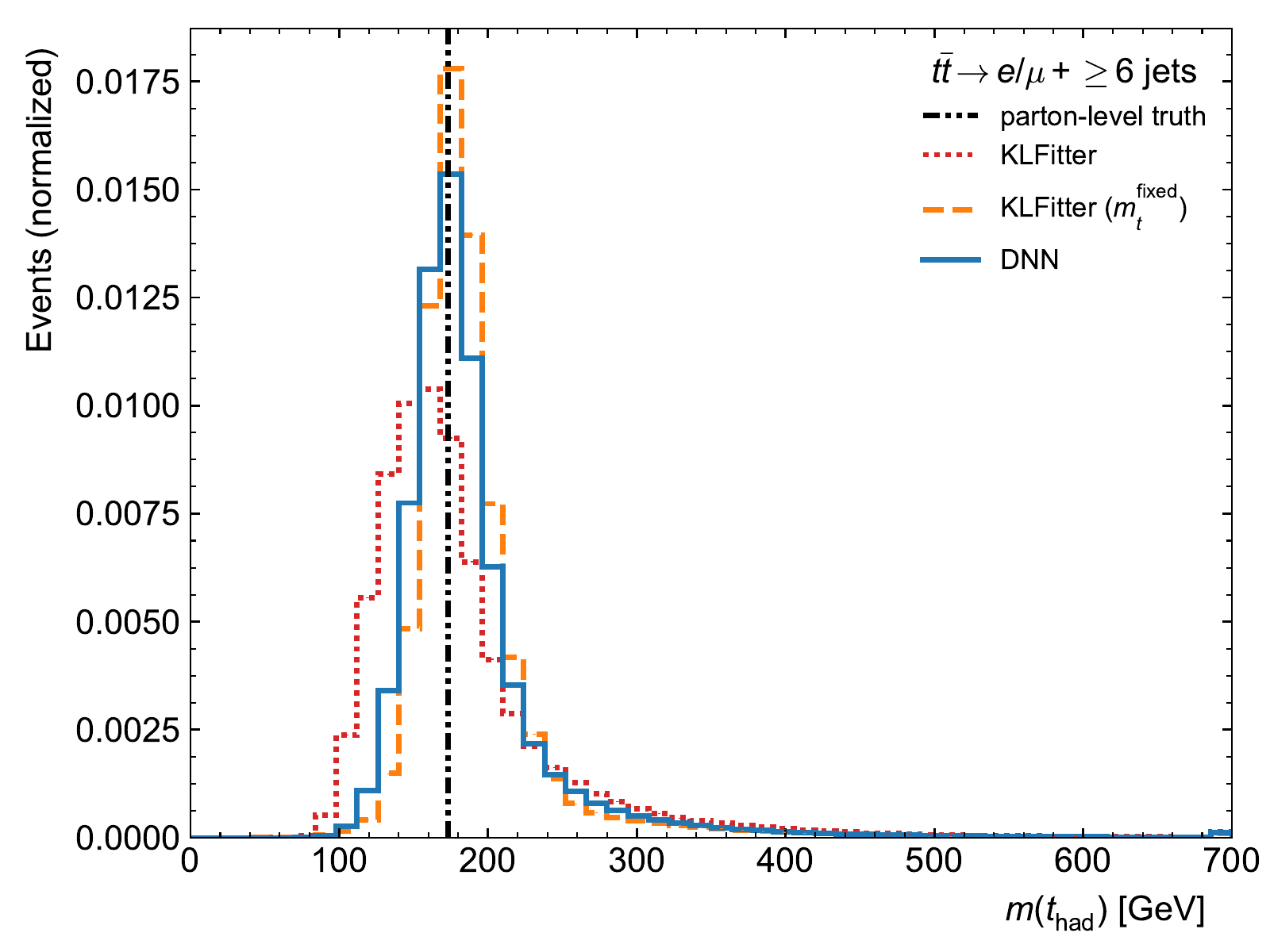}}
  \subfloat[]{\includegraphics[width=0.49\textwidth]{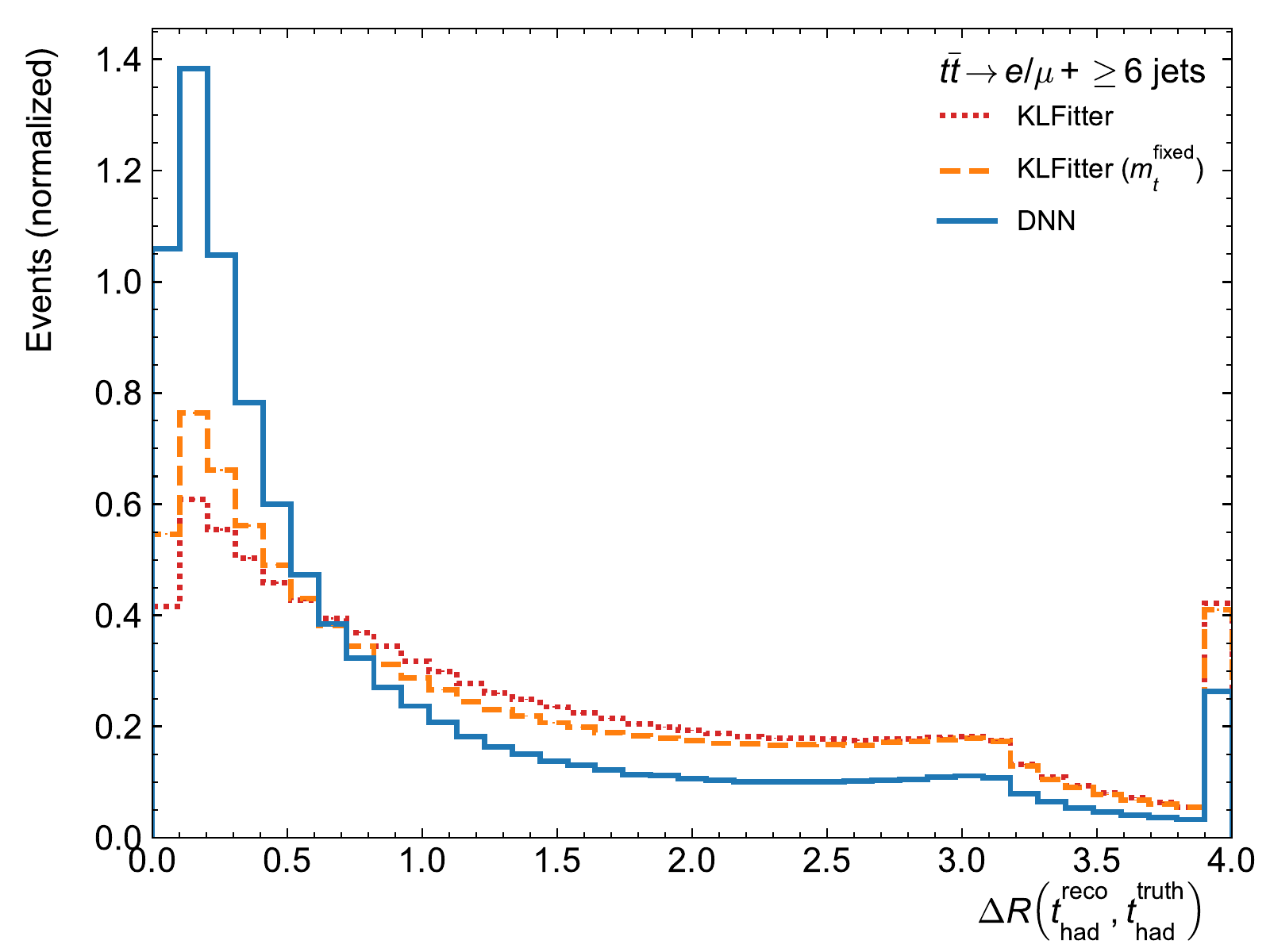}}\\
  \subfloat[]{\includegraphics[width=0.49\textwidth]{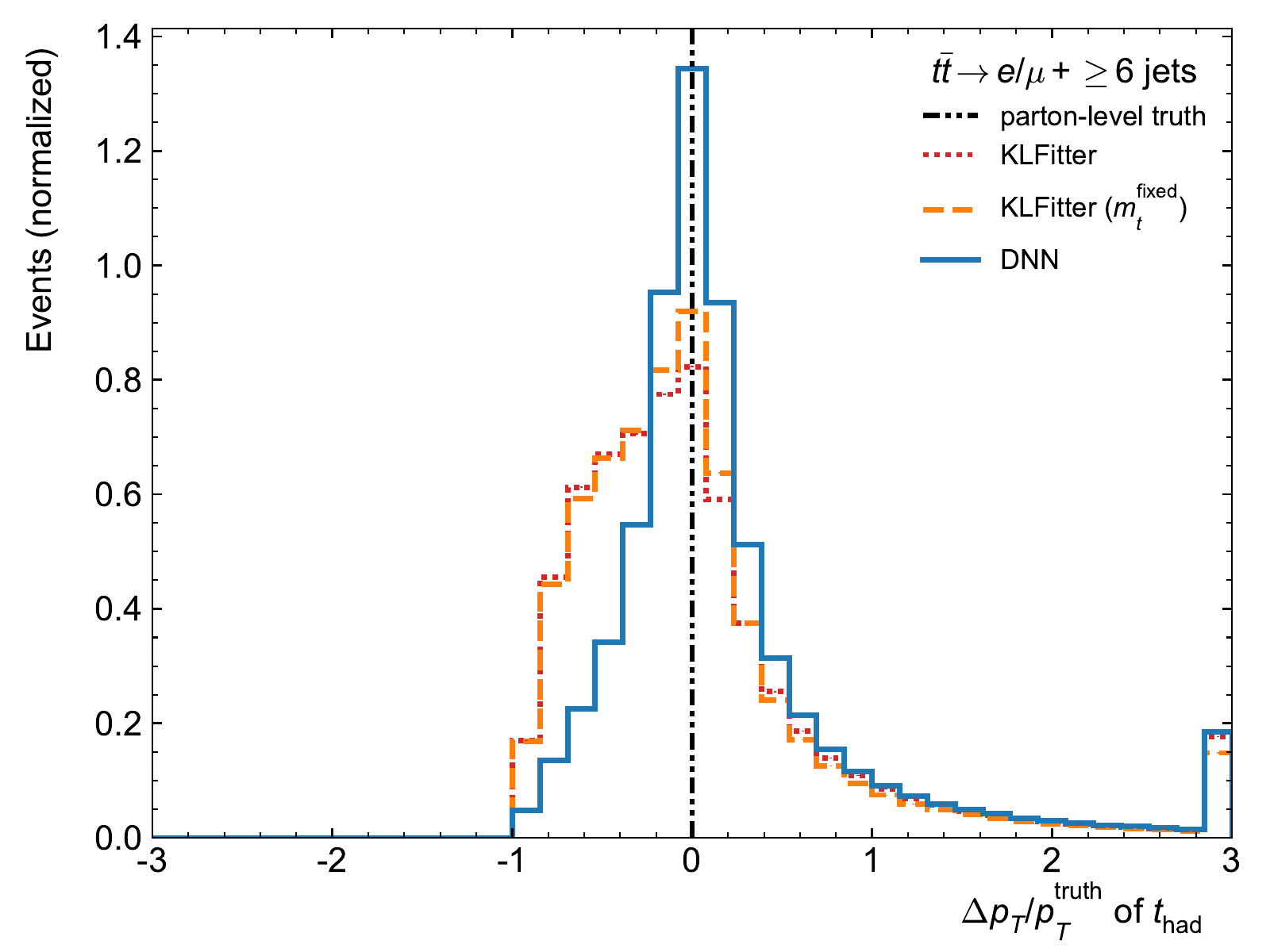}}
  \caption{Distributions for events with at least six jets of (a) the reconstructed mass of the hadronically-decaying top quark, (b) its difference in direction with respect to the true top quark in $\eta$-$\Phi$ space and (c) the relative difference of the \pt\ ($\Delta \pt/p_{\mathrm{T}}^{\mathrm{truth}}$ with $\Delta \pt = p_{\mathrm{T}}^{\mathrm{predicted}}-p_{\mathrm{T}}^{\mathrm{truth}}$) with respect to the true top-quark's \pt. In the case of the mass, the top-quark mass used in the Monte Carlo simulation is indicated by a vertical line. The performance of the reconstruction with the DNN is compared with two configurations of the \textsc{KLFitter} algorithm.}
  \label{fig:kinematics_6j}
\end{figure}

In Figures~\ref{fig:kinematics_4j}--\ref{fig:kinematics_6j}, distributions of the reconstructed mass of the top quark that decays hadronically, its difference in direction with respect to the true top quark's direction in $\eta$-$\Phi$ space and the relative difference of the \pt\ with respect to the true top-quark's \pt\ are shown for the four-, five- and six-jet selections.

The top-quark mass distribution obtained with the DNN is similar to the distribution obtained with the \textsc{KLFitter} algorithm in the fixed-mass configuration with the \textsc{KLFitter} distribution shifted to slightly larger values than the true top-quark mass (Figures~\ref{fig:kinematics_4j}(a)--\ref{fig:kinematics_6j}(a)), as also observed using only unambiguously matched events. Mean values and standard deviations of a Gaussian fit to the reconstructed mass peak in a window of $\pm 50~\GeV$ around the true top-quark mass are presented in Table~\ref{tab:resolution}, showing a similar mass resolution for the DNN and \textsc{KLFitter} in the fixed-mass mode, the slight upward shift of the \textsc{KLFitter} reconstruction in this mode, and the significantly worse resolution obtained with \textsc{KLFitter} in the floating-mass mode.

The DNN leads to a larger fraction of top quarks within $\Delta R = 1.0$ of the true top quark's direction (Figures~\ref{fig:kinematics_4j}(b)--\ref{fig:kinematics_6j}(b)). We observe again that the improvements due to the use of the DNN are stronger in the more challenging cases of events with more than four jets. The fractions of reconstructed top quarks within a radius of $\Delta R = 1.0$ of the true top quark's direction are shown in Table~\ref{tab:resolution}, quantifying the improvement that is obtained when using the reconstruction with the DNN compared to the reconstruction with \textsc{KLFitter}. Using the DNN for the reconstruction, the fractions even increase with the five- and six-jet selection compared to the four-jet selection, while for both \textsc{KLFitter} modes the opposite trend is observed. One explanation for this increase is that with the four-jet selection, one of the jets originating from the hadronically-decaying top quark is often not selected, but is chosen in the five- or six-jet selection, and the DNN is able to identify the correct permutation while the \textsc{KLFitter} reconstruction is not able to suppress the large combinatorial background. A similar behaviour is observed for the relative \pt\ resolution of the hadronically-decaying top quark (Figures~\ref{fig:kinematics_4j}(c)--\ref{fig:kinematics_6j}(c)). While only a small improvement in resolution is seen for the DNN compared to the two \textsc{KLFitter} modes for the selection with at least four jets, the improvement is much stronger in the case of the five- and six-jet selections.

\begin{table}
\centering
\caption{Mean ($m_t$) and standard deviation ($\sigma$($m_t$)) of a Gaussian fit to the peak of the reconstructed mass distribution for the hadronically-decaying top quark, and the fraction of hadronically-decaying top quarks that are reconstructed within $\Delta R = 1.0$ of the true top quark, $f\left(\Delta R < 1.0\right)$, for different jet selections and for reconstruction with the DNN and with the two configurations of \textsc{KLFitter}. The statistical uncertainties due to the limited size of the MC samples are smaller than the precision with which the values are cited.}
\begin{tabular}{llccc}
\toprule
Jet selection & Algorithm & $m_t$ & $\sigma$($m_t$) & $f\left(\Delta R < 1.0\right)$ \\
\midrule
\multirow{3}{*}{$\geq 4$ jets} & DNN & 175.1~\GeV & 25.4~\GeV & 58.7\% \\
& \textsc{KLFitter} ($m_t^{\mathrm{fixed}}$) & 178.3~\GeV & 25.5~\GeV & 55.5\% \\
& \textsc{KLFitter} & 175.1~\GeV & 27.6~\GeV & 51.5\% \\
\midrule
\multirow{3}{*}{$\geq 5$ jets} & DNN & 174.1~\GeV & 23.7~\GeV & 66.0\% \\
& \textsc{KLFitter} ($m_t^{\mathrm{fixed}}$) & 178.3~\GeV & 23.0~\GeV & 53.9\% \\
& \textsc{KLFitter} & 168.4~\GeV & 28.3~\GeV & 50.5\% \\
\midrule
\multirow{3}{*}{$\geq 6$ jets} & DNN & 172.9~\GeV & 22.8~\GeV & 66.7\% \\
& \textsc{KLFitter} ($m_t^{\mathrm{fixed}}$) & 177.8~\GeV & 20.1~\GeV & 48.3\% \\
& \textsc{KLFitter} & 161.5~\GeV & 27.5~\GeV & 44.3\% \\
\bottomrule
\end{tabular}
\label{tab:resolution}
\end{table}

Although the DNN was only trained using unambiguously matched events, the improvement in performance compared to \textsc{KLFitter} holds also when kinematic top-quark variables are reconstructed using all events that pass the selection. The fact that the DNN outperforms \textsc{KLFitter} in the reconstruction task implies that the DNN learns from additional information in the events that is not used in the \textsc{KLFitter} algorithm. As the improvement in performance is more pronounced for events with additional jets, this indicates that such information is in particular related to additional radiation, which is not part of the leading-order picture that is the basis for the \textsc{KLFitter} algorithm.

Interestingly, the DNN output is positively correlated with the \pt\ of the true hadronically-decaying top quark, with correlations of 24\%, 30\% and 32\% in the case of the four-, five- and six-jet selections. This means that a better reconstruction is achieved for a larger top-quark \pt. This is consistent with the improved performance of $\ttbar$ reconstruction found for the \textsc{KLFitter} algorithm~\cite{Erdmann:2013rxa}.

\clearpage

\section{Conclusions}
\label{sec:conclusions}
We have used a deep neural network, trained with Monte-Carlo simulated events, to reconstruct $\ttbar$ decays in the lepton+jets channel. We have optimized the hyperparameters of the neural network by maximizing the reconstruction efficiency, i.e.\ the probability to identify the correct jet permutation out of all possible jet permutations, separately for events with at least four, at least five and at least six jets. We have found that deep networks with five or six hidden layers and up to 512 nodes in the first hidden layer are necessary to reach the best performance. We have compared this approach to a widely-used kinematic fit and we have found significant improvements in identifying the correct jet permutation. We have observed particularly large improvements for the more challenging reconstruction of events with more than four jets, where at least one additional jet is selected that does not correspond to the leading-order $\ttbar$ lepton+jets topology.

While the network was trained using only events in which all considered jets were geometrically matched to the true partons from the $\ttbar$ decay, the effect of the improved reconstruction on several kinematic top-quark properties was studied using all events. While the network's performance in events with at least four jets was found to be slightly better than the benchmark algorithm, larger improvements were found for events with more than four jets.

We conclude that deep neural networks provide improvements for the reconstruction of $\ttbar$ events, which promise to improve the precision of top-quark measurements at the LHC and at the HL-LHC. We have documented our workflow for the optimisation of the hyperparameters of the deep neural network and this workflow can be followed by experimental collaborations to retrain the network taking into account their detailed detector simulations.

\section{Acknowledgements}
The authors acknowledge the support of the Funda{\c{c}}{\~{a}}o para a Ci{\^{e}}ncia e a Tecnologia (Lisbon, Portugal) through project POCI-01-0145-FEDER-029147.

\bibliographystyle{JHEP}
\bibliography{jinst}

\end{document}